\documentclass[12pt,nofootinbib,preprint]{revtex4-1}
\usepackage{amsmath,amssymb}
\usepackage{soul}
\usepackage{verbatim,graphicx}
\usepackage[usenames,dvipsnames]{color}
\usepackage[bookmarks]{hyperref}
\usepackage[all]{xy}

\newcommand{\BR}{\mathbb{R}}
\newcommand{\BZ}{\mathbb{Z}}

\newcommand{\Acal}{\mathcal{A}}

\newcommand{\Ecal}{\mathcal{E}}
\newcommand{\Hcal}{\mathcal{H}}
\newcommand{\Ncal}{\mathcal{N}}

\newcommand{\lp}{\left(}
\newcommand{\rp}{\right)}

\newcommand{\ls}{\left|}
\newcommand{\rs}{\right|}
\newcommand{\lr}{\left.}
\newcommand{\rr}{\right.}

\newcommand{\Tr}{\text{Tr}}
\newcommand{\diag}{\text{diag}}
\newcommand{\ket}[1]{\left| #1\right\rangle}

\allowdisplaybreaks

\begin{document}

\setstcolor{red}

\title{Spinning the fuzzy sphere}

\author{David Berenstein$^{\dagger,\ddagger}$, Eric Dzienkowski$^\ddagger$, Robin Lashof-Regas$^\ddagger$}
\preprint{DAMTP-2015-28}

\affiliation{$\!^\dagger$ Department of Applied Mathematics and Theoretical Physics, 
University of Cambridge, Wilberforce Road, Cambridge CB3 0WA, United Kingdom\\
$^\ddagger$ Department of Physics, University of California Santa Barbara, Santa Barbara, California 93106}

\begin{abstract}We construct various exact analytical solutions of the $SO(3)$ BMN matrix model that correspond to rotating fuzzy
 spheres and rotating fuzzy tori. These are also solutions of Yang Mills theory compactified on a sphere times time and they are also translationally invariant solutions of the $\Ncal = 1^*$ field theory with a non-trivial charge density.
The solutions we construct have a $\BZ_N$ symmetry, where $N$ is the rank of the matrices. After an appropriate ansatz, we reduce the problem to solving a set of polynomial equations
in $2N$ real variables. These equations have a discrete set of solutions for each value of the angular momentum. We study the phase structure of the solutions for various values of $N$. Also the continuum limit where $N\to \infty$, where the problem reduces to finding periodic solutions of a set of coupled differential equations. We also study the topology change transition from the sphere to the torus. 
 \end{abstract}

\maketitle

%%%%%%%%%%%%%%%%%%%%%%%%%%%%%%%%%%%%%%%%%%%%%%%%%%%%%%%%%%%%%%%%%%%%%%%%%%%%%%%

\section{Introduction}

The BMN matrix model  \cite{Berenstein:2002jq} is a massive deformation of the BFSS matrix model \cite{Banks:1996vh}. The BFSS matrix model describes the discrete lighten quantization of M-theory on flat space. The BMN matrix model analogously describes the discrete light cone quantization of M-theory on a maximally supersymmetric plane wave. These maximally supersymmetric  plane wave geometries were constructed by taking a Penrose limit of supersymmetric $AdS\times S$ spaces in \cite{Blau:2002dy}.

The BFSS matrix model results from the dimensional reduction of $\Ncal=1$ SYM from ten dimensions downs to $0+1$ dimensions and corresponds to the dynamics of D0-branes \cite{Polchinski:1995mt}. It has an $SO(9)$ symmetry of the transverse directions and a gauged $U(N)$ symmetry, where $N$ is the rank of the matrices. The BMN matrix model splits these transverse directions into two sets with different masses, so that the bosonic symmetry reduces to an $SO(3)\times SO(6)$ subgroup of $SO(9)$. The subset of the theory where only the $SO(3)$ charged scalars are excited is the $SO(3)$ BMN matrix model. The full model also results from considering an $SU(2)_L$ invariant set of configurations in Yang-Mills theory on an $S^3\times \BR$ geometry \cite{Kim:2003rza}. Thus, any classical solution of the $SO(3)$ BMN matrix model is also a classical solution of Yang Mills theory on a sphere times time. One can also show that any solution of the BMN matrix model corresponds to a classical solution of the $\Ncal = 1^*$ field theory ( see \cite{Polchinski:2000uf} and references therein).

It is known that for generic initial conditions, the BMN matrix model is chaotic \cite{Asplund:2011qj,Asplund:2012tg,Asano:2015eha}. This can be understood from the chaotic dynamics of dimensionally reduced Yang-Mills \cite{Sav,Chirikov:1981cm,Aref'eva:1997es,Matinyan:1981ys}. However, special initial conditions can in principle be soluble analytically. 
It is expected quite generally that solutions that minimize the energy with an additional conserved quantity turned on and constrained can be stationary. This is usually handled with a Routhian if the conjugate variable to the conserved quantity can be separated.
Many of these states that minimize the energy given some conserved central charge have interpretations in terms of BPS states in supersymmetric field theories.

As it turns out, the BMN matrix model has exact, supersymmetric solutions with zero energy \cite{Berenstein:2002jq}. These matrix configurations are characterized by all adjoint representations of $\mathfrak{su}(2)$. They have an interpretation as giant gravitons \cite{McGreevy:2000cw}. The spectrum of fluctuations around these solutions is known \cite{Dasgupta:2002hx} (see also \cite{Berenstein:2010bi} for an alternative derivation of the spectrum) and one can argue that there is a large tower of protected states that are available to study \cite{Dasgupta:2002ru}. Unfortunately the nonlinear structure of the classical solutions that make this tower of BPS states is not known.

It is expected that adding angular momentum to the fuzzy sphere states can induce topology changes from a sphere to a torus \cite{Nishioka:2008ib}. Our purpose in this paper is to investigate this topology transition with a special family of matrix solutions at finite angular momentum. The paper is mostly devoted to constructing these solutions.
Once the solutions are found, the geometry of the corresponding fuzzy membrane is analyzed using the techniques in \cite{Berenstein:2012ts}. 

%%%%%%%%%%%%%%%%%%%%%%%%%%%%%%%%%%%%%%%%%%%%%%%%%%%%%%%%%%%%%%%%%%%%%%%%%%%%%%%

\section{The Hamiltonian and the Ansatz}

The $SO(3)$ BMN matrix model is a dynamical system with three $N\times N$ Hermitian matrices $X^{1,2,3}$, or alternatively $X,Y,Z$. The conjugate momentum matrices are $P_{1,2,3}$, and $P_{X,Y,Z}$ respectively.
The BMN Hamiltonian is given by
\begin{equation}
\label{eq:BMNHam}
H = \frac{1}{2}\Tr(P_1^2+P_2^2+P_3^2) + \frac{1}{2}\Tr\lp \sum_{j=1}^3 (X^j + i\epsilon_{jmn} X^m X^n)^2 \rp
\end{equation}
The system possesses a $U(N)$ gauge symmetry where $X^i$ and $P_i$ both transform in the adjoint, $X^i \to U X^i U^{-1}$ and $P_i \to U P_i U^{-1}$. The presentation of the Hamiltonian \eqref{eq:BMNHam} is in the gauge $A_0=0$. 
The generators of gauge transformations are the matrix of functions on phase space given by
\begin{equation}
\label{eq:gauss}
{\mathfrak G} = \sum_{j=1}^3 [P_j, X^j]
\end{equation}
The dynamics need to be supplemented by the Gauss' law constraint ${\mathfrak G} = 0$.
The system also enjoys an $SO(3)$ symmetry of rotations of the matrices $X,~Y,~Z$ into each other. The generator of angular momentum along the $Z$ direction is
\begin{equation}
\label{eq:Lz}
J = L_Z = \Tr(X P_Y - YP_X)
\end{equation}
with similar expressions for the other two $SO(3)$ generators.
Lastly, the equations of motion are
\begin{align}
\label{eq:EOMx}
\dot{X}^j &= \frac{\partial H}{\partial P^j} = P_j, \\
\label{eq:EOMp}
\dot{P}_j &= -\frac{\partial H}{\partial X^j} = -X^j - 3i\epsilon^{jmn}X^mX^n - [[X^j, X^m], X^m]
\end{align}

The solutions with $H = 0$ are given by fuzzy spheres. These are solutions of the equations
\begin{equation}
%X^1 + i[X^2, X^3] = 0
[X^i, X^j] = i\epsilon^{ijk}X^k
\end{equation}
The solutions to these equations are characterized by direct sums of the adjoint matrices for irreducible representations of $\mathfrak{su}(2)$.
For these solutions we have $P_1,~P_2,~P_3=0$ and thus are classically gauge invariant according to \eqref{eq:gauss}.
These solutions carry no angular momentum as $\vec{L} = 0$ identically.

The Hamiltonian also satisfies a BPS inequality bound, where $H\geq |J|$ (details can be found in \cite{Hoppe:2007tv}). Solutions that saturate the bound will be called extremal or BPS. 
This follows from writing the Hamiltonian as a sum of squares in a slightly different way
\begin{align}
H &= \Tr\lp \frac{1}{2} P_3^2 + \frac{1}{2}(P_1 \pm (X_2 + i\epsilon_{231}[X^3,X^1]))^2 \rr \nonumber \\
&\hphantom{=\Tr\lp\frac{1}{2}\rr} \lr + \frac{1}{2}(P_2 \mp (X_1 + i\epsilon_{123}[X^2,X^3]))^2 + \frac{1}{2}(X_3 + i\epsilon_{312}[X^1,X^2])^2 \rp \pm J\label{eq:Hsq}
\end{align}
The cross terms between $X_2,~P_1$ and $X_1,~P_2$ in the squares generate a copy of $J$ that needs to be subtracted. The cross terms with $P_1$ and $[X^3,X^1]$ lead to something that does not automatically cancel for generic matrices, but after a bit of reshuffling can be shown to be proportional to 
\begin{equation}
X^3 ([X^1,P^1]+[X^2,P^2])
\end{equation}
and we recognize the Gauss' law constraint starting to arise. After imposing the full Gauss' law constraint, we get $\Tr(X^3[X^3,P^3])$ that does vanish identically.

The BPS bound is not directly related to supersymmetry. Instead it is derived from the conformal group in four dimensions, where one can show based on unitarity arguments that $E\geq J$ from requiring that $K=P^\dagger$, where $P,~K$ are the generators of translations and special conformal transformations on $S^3\times \BR$. This bound asserts that the dimension of operators is greater than the spin and is usually saturated for free theories, or nearly free theories at leading order in perturbation theory. This should be lifted in general theories (see \cite{Hellerman:2015nra} for a recent discussion). However, the bound does show up in studying supersymmetric BPS states \cite{Bak:2002rq,Bak:2005ef} for the full BMN matrix model.
This bound descends to any classical solution of Yang-Mills theory which is a conformal field theory at the classical level.

Now we move on to construct solutions to \eqref{eq:EOMx} and \eqref{eq:EOMp} with non-zero $J$. We define the matrices
\begin{align}
X^+ = \frac{1}{2}(X + iY),& \quad X^- = \frac{1}{2}(X - iY) \\
X = X^+ + X^-,& \quad Y= -i(X^+ - X^-)
\end{align}
We will make an ansatz for a solution more general than the fuzzy spheres
\begin{equation}
\label{eq:ansatz}
X^+(t) = %
\begin{pmatrix}
0 & a_1 \exp(i \omega_1 t) & 0 & \dots \\
0 & 0 & a_2 \exp(i \omega_2 t) & \dots \\
\ddots & \ddots & \ddots & \vdots \\
0 & \dots & 0 & a_{N-1}\exp(i \omega_{N-1} t) \\
a_N \exp(i\omega_{N}t) & 0 & \dots & 0
\end{pmatrix}
\end{equation}
with $a_i$ constants, and $X^-(t) = (X^+(t))^\dagger$ (this is the transpose complex conjugate).
At the same time we also take
\begin{equation}
\label{eq:ansatzZ}
Z(t) = \diag(z_1, \dots, z_N)
\end{equation}
independent of time and real. We will explain the origin of this ansatz in the following section. Notice that all fuzzy sphere ground states are solutions of this kind already, with $\omega_i=0$, and some of the $a_i=0$. Gauge transformations that commute with $Z$ allow us to vary the relative phases of the $a_i$. Therefore we can assume that they are real, and a common phase can be translated away by choosing the starting time appropriately.
This ansatz is different to those that have been studied before \cite{Hoppe:2007tv} (previous works also look for solutions of the BMN and BFSS matrix model where more than three matrices are oscillating \cite{Arnlind:2003nh,Arnlind:2003nh,Arnlind:2004br}). Other time dependent solutions can be found in \cite{Steinacker:2011wb,Polychronakos:2013fma,teinacker:2014eua}, but again, these involve more matrices being turned on.  In particular, we allow for multiple frequencies to arise in the ansatz, rather than just one.
This does not affect fact that the system is rigidly rotating. One can use a gauge transformation to make the frequencies the same, but one pays the price that $A_0$, the connection in the time direction, becomes non-trivial. This is actually very useful for the BPS states, where $A_0\propto Z$ \cite{Bak:2002rq}.

The method of solving the equations is then to first solve for the conjugate momenta by using equation \eqref{eq:EOMx}. Then, we solve the Gauss' law constraint \eqref{eq:gauss} to relate the $\omega_i$ to each other. These can then be substituted into the angular momentum equation \eqref{eq:Lz}, so that we can express the $\omega_i$ in terms of the total angular momentum $L_z=J$ and the $a_i$. One can then show that the system of equations of motion  \eqref{eq:EOMp} reduces consistently to an algebraic set of real equations for the $a_i,~z_i$ and that it has as many unknowns as there are variables. One therefore expects to generally find a discrete (possibly empty) set of solutions to the equations. We will show eventually that this set of solutions is non-empty for all $J$.

%%%%%%%%%%%%%%%%%%%%%%%%%%%%%%%%%%%%%%%%%%%%%%%%%%%%%%%%%%%%%%%%%%%%%%%%%%%%%%%

%\subsubsection{Relation to SYM and Polchinski-Strassler}
\subsection{Relation to SYM and $\Ncal=1^*$}

The BMN hamiltonian also arises from sphere reductions of four dimensional Yang-Mills into a $S^3\times \BR$ \cite{Kim:2003rza} and from the $\Ncal=1^*$ theory. The relation to Yang Mills on $S^3\times \BR$ is as follows. Consider that the round three sphere is also the group manifold of $SU(2)$, and has an $SU(2)\times SU(2)$ isometry group by acting with the group on the left and on the right. We can use a basis of left invariant one-forms under $SU(2)$, $e^{1,2,3}$ and write the spatial part of the connection connection as follows
\begin{equation}
A = A_i e^i
\end{equation}
where the $A_i$ are now Lie-algebra valued functions on $S^3$. Requiring that the allowed configurations are invariant under left actions of the group, we have that $A_i$ becomes position independent and is just a constant hermitian matrix. The Mauer-Cartan equations for the one forms $e^i$ then give us that 
\begin{equation}
dA + A\wedge A = A_i d e^i + [A_i, A_j] e^i \wedge e^j
\end{equation}
whereas the electric fields will give
\begin{equation}
D_t A \simeq (D_t A_i) dt \wedge e^i
\end{equation}
Because of the large amount of symmetry preserved, we are led to a consistent truncation of the SYM lagrangian. After plugging in, we identify $A_i\simeq X_i$ and $P^i \simeq D_t A_i$. The Legendre transform of the Lagrangian for $SU(2)$ invariant fields will give rise to the same BMN Hamiltonian.

For the $\Ncal=1^*$ field theories, they are obtained from $\Ncal = 4$ SYM after an $SO(3)$ invariant mass deformation of the super-potential. The $\Ncal = 4$ SYM Lagrangian can be characterized by having three chiral matter super fields $\phi^{1,2,3}$ in the adjoint of $SU(N)$, and a super potential of the form
\begin{equation}
W = \Tr(\phi^1\phi^2\phi^3 - \phi^3\phi^2\phi^1)
\end{equation} 
and the full potential has an $SO(6)$ R-symmetry invariance, of which only an $SU(3)$ rotation is apparent in terms of $\Ncal = 1$ super fields. 
 
The${\Ncal}=1^*$ deformation adds the super potential mass term
\begin{equation}
\delta W = -\frac{i}{2} M \Tr\lp \sum_{k=1}^3 (\phi^k)^2\rp
\end{equation}

The factor of $i$ is a choice of convention, as the phase of $M$ can be changed by a global R-charge rotation of the $\phi$. This preserves an $SO(3)$ global symmetry from the R-charge, but the theory can confine in the infrared in some of it's vacua. For this theory, the classical vacua are given by fuzzy spheres, or $\mathfrak{su}(2)$ representations \cite{Vafa:1994tf} (see also \cite{Dorey:1999sj} for the characterization of the vacua at strong coupling).
The potential for the $\phi$ fields is given by
\begin{equation}
V(\phi) = \sum_{k=1}^3 \ls [\phi^\ell,\phi^m] \epsilon_{\ell m }\!^{k}+i M \phi^k \rs^2 + \lp\sum_{k=1}^3 [\phi^k,\bar\phi^k]\rp^2 
\end{equation}
where the first term is the F-term and the last term comes from the D-terms of YM.
The theory also has a parity transformation that sends the super field $\phi^i\to \bar \phi^i$. This is a symmetry of $\Ncal = 4$ SYM if the theta angle of the field theory vanishes. 
The scalar potential is invariant under this transformation. We can choose to look for configurations that are classically parity invariant. In that case, we only preserve the real part of $\phi^i$, while the imaginary part is removed. This parity transformation commutes with the $SO(3)$ symmetry group, but not the full $SU(3)$.

With this constraint on the fields, the potential term arising from the D-terms automatically vanishes. Moreover, the potential for $V(\phi)$ becomes the BMN potential after factoring out the dimensionful constant from the fields $\phi$. A translation invariant classical solution of the $\Ncal=1^*$ field theory with  parity invariance can be understood as a classical solution of the $SO(3)$ BMN matrix model. The angular momentum of the BMN matrix model solutions becomes a charge density for a global symmetry of the field theory. This is a charge density in a four dimensional $\Ncal = 1$ field theory, so it is not a charge density for a central charge. The solutions are to be regarded as non-supersymmetric, but in the classical limit they are controlled by the same dynamics as the $SO(3)$ BMN matrix model. The quantum corrections will be different. The structure of solutions will then be described by the (parity invariant) phase diagram for the weakly coupled $\Ncal=1^*$ field theory at finite charge density. This in turn can be understood as a phase diagram for a (top-down) holographic superconductor \cite{Hartnoll:2008vx}.

%%%%%%%%%%%%%%%%%%%%%%%%%%%%%%%%%%%%%%%%%%%%%%%%%%%%%%%%%%%%%%%%%%%%%%%%%%%%%%%

\section{Symmetry Considerations}

A rather natural question to ask is if we can find rotationally invariant configurations of the $SO(3)$ BMN matrix model around the $Z$ axis that rotate uniformly without changing the shape of the configuration and are a small perturbation of a single fuzzy sphere configuration. After all, this is how the giant torus configurations in supergravity are constructed \cite{Nishioka:2008ib}. It turns out that the answer is no. 

The way to see this is the following. Assume that $Z$ is a Hermitian matrix with eigenvalues that are not degenerate. A configuration will be rotationally invariant around the $Z$-axis if a naive rotation of the matrices into each other can be undone with a gauge transformation. This is the same way that the method of images works for D-branes on orbifolds when considering a discrete subset of the rotation group \cite{Douglas:1996sw} (this is also the mechanism for rotational invariance of monopole solutions in nonabelian gauge theories \cite{'tHooft:1974qc,Polyakov:1974ek}).

That is, if for any angle $\theta$ we can find a unitary matrix $U(\theta)$ such that 
\begin{align}
U(\theta) Z U^{-1}(\theta) &= Z \nonumber\\
\label{eq:rotinv}
U(\theta) X U^{-1}(\theta) &= X \cos(\theta) - Y\sin(\theta) \\
U(\theta) Y U^{-1}(\theta) &= Y \cos(\theta) + X\sin(\theta)  \nonumber 
\end{align}
Because $Z$ has non-degenerate eigenvalues, the first equation tells us that $U$ must be diagonal in the same basis that $Z$ is (they commute with each other). We gauge transform to such a basis without loss of generality. The angle $\theta$ will be identified with the time evolution itself later on.
The uniform rotation motion of the configuration requires that $Z$ is time independent, so that $\dot{Z} = 0$.

Let $\ket{1}, \dots, \ket{N}$ denote the eigenvectors of $Z$. It is also convenient to rewrite the last two equations of \eqref{eq:rotinv} in terms of the general matrix $X^+ = X+iY$ and it's adjoint. 
The matrix $X^+$ is unconstrained. Then we have that
\begin{equation}
\label{eq:inv1}
U(\theta) X^+ U^{-1}(\theta) = \exp(i\theta) X^+
\end{equation}
Using a general expression for $X^+\ket m = X^+_{mn} \ket{n}$, and $U\simeq \diag(\exp(i \theta_i))$ we find that $X^+_{mn}$ transforms under conjugation by $U$ as 
\begin{equation}
X^+_{mn}\to \exp(i(\theta_m - \theta_n)) X^+_{mn}
\end{equation}
which can only be equal to $\exp(i\theta) X_{mn}$ if $\theta_m-\theta_n= \theta \mod(2\pi)$ or if $X^+_{mn}=0$. Picking $\theta$ arbitrarily small, we can only satisfy 
$\theta_m-\theta_n=\theta$ for some of the components, and because $\theta$ is very small, we can arrange the kets $\ket{n}$ such that the $\theta_n$ are strictly decreasing as we increase $n$. This shows that $U(1)$ invariance requires $X^+_{mn}$ to be upper triangular with zeros on the diagonal. Now, comparing with the fuzzy sphere solutions, $X^+$ actually must border the diagonal. That is, we find that the solutions must be of the form
\begin{equation}
\label{eq:ansatzrotinv}
X^+ =
\begin{pmatrix} 0 & |a_1|\exp(i\phi_1) & 0& \dots \\
0 & 0 & |a_2| \exp(i\phi_2) & \dots\\
\ddots & \ddots & \ddots & \vdots \\
0 & \dots & 0 & |a_{N-1}|\exp(i\phi_{N-1}) \\
0 & 0 & \dots & 0
\end{pmatrix}
\end{equation}
where the $a_i$ are some (as of yet unspecified numbers, that depend on time), and both $X^+ X^-$ and $X^- X^+$ are diagonal and rotationally invariant themselves. Thus if a configuration is rotating uniformly we find that the only possible solution has the $|a_i|$ being constant, and all the dynamics will be in the phases $\phi_i(t)$. Inserting these expressions in the Gauss' law constraints, we find that 
\begin{equation}
|a_i|^2 \dot{\phi}_i-|a_{i+1}|^2 \dot{\phi}_{i+1} = 0
\end{equation} 
for all $i$. This leads us to show that $\dot{\phi}_{N-1}=0$ and from there, $\dot{\phi}_i=0$ for all $i$. The solutions are thus static and carry no angular momentum. Since we want to turn on angular momentum, we need to relax the rotational invariance around the $Z$-axis of the configuration.

The obvious idea is to look for the maximal discrete subgroup of rotations that can actually be preserved if we can not have a full $SO(2)$ symmetry. Let us assume that we turn on an $X_{mn}$ which is not one of the above. The matrix $U$ that implements the constraint \eqref{eq:inv1} for $X^+$ defined in \eqref{eq:ansatzrotinv} is diagonal and up to a global phase it is equal to
\begin{equation}
U = \diag[\exp(-i\theta), \exp(-2i\theta), \exp(-3i \theta)\dots \exp(-N i \theta)]
\end{equation}

We then have that $n\theta -m\theta= \theta \mod(2\pi)$, or equivalently, that $(m+1-n)\theta= 0 \mod(2\pi)$. This tells us that we can preserve a $\BZ_{m+1-n}$ subgroup of the $SO(2)$ rotations if we turn on this particular $X_{mn}$. If we want to maximize this group, we find that we must take $m=N$ and $n=1$, where we get an $\BZ_N$ subgroup of the rotation group to be invariant. None of the other $X_{m n}$ that are not already turned on are neutral under this subgroup. We can self-consistently set them to zero by symmetry arguments; if an initial solution for $X,~P$ respects the symmetry, the dynamics being $\BZ_N$ invariant will guarantee that no symmetry breaking can occur afterwards. We are therefore led to a general ansatz for $X^+$ which can still have arbitrary time dependence
\begin{equation}
X^+ = %
\begin{pmatrix} 0& |a_1|\exp(i\phi_1) & 0 & \dots & 0 \\
0 & 0 & |a_2| \exp(i\phi_2) & \dots & 0 \\
\vdots & \vdots & \ddots & \ddots & \vdots\\
0 & \ddots & \dots & 0 & |a_{N-1}|\exp(i\phi_{N-1}) \\
|a_N|\exp(i\phi_N) & 0 & \dots & \dots & 0
\end{pmatrix}\label{eq:ansatznew}
\end{equation}

Now we want to insist that the only motion that the system does is rigid rotation, so that the $|a_i|$ are necessarily constant, but the $\phi_i$ can vary in time. Substituting this in Gauss' law
shows that all the motions in the angles are related to each other by $|a_i|^2\dot{\phi}_i -|a_{i+1}|^2 \dot{\phi}_{i+1}=0$ cyclically. Furthermore, one can show that $L_z=\frac{1}{2} \sum_i |a_i|^2 \dot{\phi}_i=J$ which is a conserved quantity. Putting these two pieces of information together shows that the phases $\phi_i$ have constant time derivatives which we call $\omega_i$. This leads us to the general form of the ansatz described in equations \eqref{eq:ansatz} and \eqref{eq:ansatzZ}. 

Notice also that the ansatz we have made has an additional remnant discrete $\hat\BZ_N$ symmetry on the variables $a_i,~z_i$, where we send $|a_k|\exp(i\phi_k)\to |a_{k+1}| \exp(i\phi_{k+1})$ and $z_k\to z_{k+1}$  cyclically. That is, any solution of the equations can be permuted to a new solution using this symmetry. This is a subset of the gauge freedom of the original system, where we permute the eigenvalues of a hermitian matrix. In D0 brane dynamics this permutation symmetry is associated to the permutation statistics of D-branes \cite{Banks:1996vh}. This symmetry can also be understood if we orbifold the matrix problem by the original $\BZ_N$ symmetry we identified  via the rules of \cite{Douglas:1996sw}. In the orbifold theory by an abelian symmetry, it is expected that one has a
dual quantum symmetry $\hat \BZ_N$ that permutes the nodes of the corresponding quiver theory \cite{Berenstein:2000hy}.  Gauging this dual quantum symmetry restores the original theory in a straightforward way \cite{Berenstein:2001jr}. There is a second symmetry where we reverse the order of the $|a_i|,~z_i$ and also change the sign of the $z_i$.
This acts essentially as reflection on the $Z$-axis, so that when we combine it with time reversal, we can still spin in the same direction. 
These two symmetries form a dihedral group with $2N$ elements and it will be useful for analyzing the set of solutions of the ansatz. 

Notice that if a discrete subgroup of the quantum symmetry is left unbroken, this implies that there is an enhanced unbroken gauge group so long as it is not just the spatial reflection symmetry. For example, if the $\hat{Z}_N$ is unbroken (we find some solutions of this type), the unbroken gauge group turns out to be $U(1)^N$. The solution can be interpreted as $N$ D0 branes separated from each other in a $\BZ_N$ symmetric pattern of rotations around the origin, just like one would expect from the method of images in an orbifold.

%%%%%%%%%%%%%%%%%%%%%%%%%%%%%%%%%%%%%%%%%%%%%%%%%%%%%%%%%%%%%%%%%%%%%%%%%%%%%%%

\section{The Solutions of the Ansatz are a Set of Critical Points of an Energy Function}

The equations of motion that follow from the Hamiltonian \eqref{eq:BMNHam} come in two different sets. First, we have that 
\begin{equation}
\dot{X}^j = P_j
\end{equation}
We can solve these immediately given the ansatz for the $X$. We find that $P_Z=0$ and that
\begin{align}
P_X = \dot{X} &= i \Omega X^+ - iX^- \Omega \\
P_Y = \dot{Y} &= \Omega X^+ + X^-\Omega
\end{align}
where the matrix $\Omega$ of angular velocities is given by
\begin{equation}
\Omega = \diag(\omega_1,\omega_2,\dots,\omega_N)
\end{equation}
This way we find that
\begin{align}
P_X X - X P_X &= (i\Omega X^+ - iX^-\Omega)(X^+ + X^-) - (X^+ + X^-)(i \Omega  X^+ - i X^- \Omega)\\
P_Y Y - Y P_Y &= (\Omega X^+ + X^-\Omega)(-i(X^ + -X^-)) - (-i(X^+-X^-)) (\Omega X^+ + X^-\Omega)
\end{align}
Adding these two, we find that the terms with two $X^+$ or two $X^-$ cancel each other, so that the Gauss' law constraint reads
\begin{equation}
Q = 2i\Omega X^+ X^- - 4i X^- \Omega X^+ + 2iX^+X^-\Omega 
\end{equation}
A straightforward computation shows that this is a diagonal matrix and that it is equal to
\begin{equation}
\label{eq:gaussconst}
Q = 4i\diag(|a_i|^2\omega_i -|a_{i+1}|^2 \omega_{i+1})
\end{equation}
To satisfy the Gauss law constraint, we need that $Q=0$ identically. Hence we find that $|a_i|^2 \omega_i$ is independent of $i$.

Similarly, we find that the angular momentum is written as
\begin{equation}
J = \Tr((X^+ + X^-)(\Omega X^+ + X^-\Omega) - (-iX^+ + iX^-)(i\Omega X^+ - iX^- \Omega))
\end{equation}
and similarly, the terms with two copies of $X^+$ or $X^-$ cancel even before taking the trace.
We find that 
\begin{equation}
J= 2\Tr(X^+ X^- \Omega + X^-\Omega X^+) = 2\Tr(\diag(\omega_i|a_i|^2 + \omega_{i-1} |a_{i-1}|^2))
\end{equation}
Now, notice that because of Gauss' law constraint, all the $\omega_i |a_i|^2$ are equal to each other. Thus, even before taking the trace, the matrix version of $J$ is proportional to the identity. We can interpret this as having uniform density of angular momentum per unit D0-brane of the corresponding fuzzy membrane.
We use this to find that
\begin{equation}
J = 4 N \omega_i |a_i|^2
\end{equation}
so that we can substitute
\begin{equation}
\omega_i = \frac{J}{4N|a_i|^2}
\end{equation}
Knowing the $|a_i|$, the $z_i$ and $J$, we can evaluate the energy by substituting the above results in the Hamiltonian \eqref{eq:BMNHam}. The result for the kinetic energy is
\begin{align}
E_{kin} (J, |a_i|) &= \frac{1}{2} \Tr(P_X^2+P_Y^2) = \frac{1}{2}\Tr (P_X + iP_Y)(P_X - iP_Y) \\
&= \frac{4}{2} \Tr(\Omega X^+  X^- \Omega)\\
&= 2\sum_{i=1}^N |a_i|^2 \omega_i^2 = \frac 1{8N^2 }\sum_{i=1}^N \frac{J^2}{|a_i|^2}
\end{align}
What is important to realize is that this looks very similar to an angular momentum centrifugal potential, where each particle (associated to the radial variable $a_i$) has the same angular momentum $J/N$ and the same mass (in this case the mass would be interpreted as $4$). 
To specify the full problem, we need to evaluate the potential energy as well. A straightforward, though rather tedious procedure shows that
\begin{equation}
V(|a_i|,z_i) = \sum_{i=1}^N \frac{1}{2} [z_i+2|a_{i-1}|^2-2 |a_i|^2]^2+2(1+z_{i+1}-z_i)^2 |a_i|^2\label{eq:nn}
\end{equation}
where the $i$ are defined modulo $N$.
We denote the full energy of a configuration $(J,~|a_i|,~z_i)$ by 
\begin{equation}
\Ecal(J,|a_i|,z_i) = E_{kin}(J, |a_i|) + V(|a_i|, z_i)
\end{equation}

When we consider the energy function $\Ecal(J, |a_i|,z_i)$ and recall that $P_Z=0$ in our ansatz, it is straightforward to notice that the equations of motion of the $Z$ variables are exactly the equations that extremize $\Ecal$ as a function of the $z_i$ keeping the other variables fixed. Namely, that $\partial_{z_i}\Ecal = 0$. This suggests that to look for solutions of the original problem we set out to do with our ansatz, it is enough to consider the extrema of the $\Ecal$ energy function at fixed $J$. Indeed, any solution of the ansatz that solves the equations of motion of the Hamiltonian \eqref{eq:BMNHam} are going to be extrema of the energy function $\Ecal$ and vice versa, any extremum of the energy function can be shown to give a solution of the equations of motion derived from the $SO(3)$ BMN Hamiltonian.

The essence of the proof is that when we take the derivatives of the kinetic energy with respect to the $|a_i|$ we find that 
\begin{equation}
\label{eq:momentumdot}
\partial_{|a_i|} E_{kin} = -\frac{J^2}{4 N^2|a_i|^3} = -4\omega_i^2 |a_i|
\end{equation}
and these can be assembled into $\dot{P}_X = \ddot{X}, \dot{P}_Y = \ddot{Y}$.
Then we need to compare this expression to the derivatives of the potential
\begin{equation}
\label{eq:force}
\partial_{|a_i|} V
\end{equation}
which can be assembled into the right hand side of the Hamilton's equations for \eqref{eq:BMNHam}. The two sets of equations can be shown to be the same set when we remove the time dependent phases $\exp(i \omega_i t)$.

The potential is a sum of `nearest neighbor' terms and it is a sum of squares. As a function of the $z_i$ it is quadratic. Therefore if we fix the $|a_i|$, we can solve for the $z_i$ via a linear set of equations, and these equations are independent of $J$. The quadratic form that appears in front of the $z_i$ is of the form $\delta_{ij} + U_{ij}$ where $U$ is a non-negative matrix. Therefore it is always invertible.

The kinetic energy is also a sum of squares. Therefore the energy function is bounded from below.
Notice that when $J\neq 0$, and as we take $|a_i|\to 0$ the energy diverges at $|a_i|=0$. Also, when we take the $|a_i|\to \infty$ the potential grows quadratically in the $|a_i|$, and if we solve for the $z_i$ in this limit, they are bounded. Therefore we expect that generically the potential grows quartically at infinity. The only time when this does not happen is when we take a double scaling limit where $|a_{i}|^2 = \Lambda \to \infty$ independent of $i$. In this case one has that the $z_i\to 0$ and the potential grows only like $\Lambda^2$. 

Considering the configuration space as the set of open intervals $|a_i|\in (0,\infty)$ (since after all we can solve for the $z_i$ given the $a_i$), we have that the domain of interest is an open ball (it is diffeomorphic to $(0,1)^N$), and the potential function diverges on all the boundary, while it is finite (and indeed analytic) in the interior. This shows that the energy function has at least one minimum. Moreover, if we compactify the boundary of the configuration space by adding one point at infinity, we get a configuration space which is  compact and has the topology of a sphere. The point at infinity realizes the maximum of the energy function continuously \footnote{We can always map the energy function from $[0,\infty)$ to $[0,1]$ by using $\tanh(\Ecal)$ for example.}.

Notice that 
\begin{equation}
\partial_{|a_i|}\Ecal_J(|a_i|)= \partial_{|a_i|}\Ecal(J,z_i(a_i), a_i)|_z+ \partial_{z_j} \Ecal(J,z_i(a_i),a_i) \frac{\partial z_j}{\partial |a_i|}= \partial_{|a_i|}\Ecal(J,z_i(a_i),a_i)|_z
\end{equation}
since the $z_i$ solve the $ \partial_{z_j}\Ecal(J,z_i(a_i),a_i) =0$ equations. Any critical point of the original $\Ecal(J,a_i, z_i)$ will give rise to a critical point of $\Ecal_J(|a_i|)$ and vice versa. In our original ansatz where we have $3N$ variables, given by $a_i, \omega_i, z_i$, we have managed to reduce the problem to a set of algebraic  equations in $N$ variables, the $a_i$ themselves. 

For the purpose of analysis of configurations, the function $\Ecal_J(|a_i|)$ where we fix the $J$ and have already solved for the $z_i$ will be thought of as a Morse function on this topological sphere (the reader unfamiliar with Morse theory should look at \cite{Witten:1982im,matsumoto}, and the lecture notes by Hutchins are very approachable \cite{Hutchinslecnotes}). The main reason for  using Morse theory is that the main ingredient that it utilizes and analyzes is the set of critical points of a function, and from this set the topology of the manifold can be reconstructed. For us, the topology is already known (it is a sphere), but the set of critical points is not. The set of critical points and gradient flows between them gives a model for the cohomology of the manifold, which in our case is known. Given a set of critical points we can ask if the set is consistent with the topology of the manifold. If it is not, we are missing critical points. Also, as $J$ changes, the set of critical points can change dimension and how these changes can happen is understood in general.  In our case, because of the extra symmetries of the potential, the critical points will exhibit also a representation of the symmetry group.

%%%%%%%%%%%%%%%%%%%%%%%%%%%%%%%%%%%%%%%%%%%%%%%%%%%%%%%%%%%%%%%%%%%%%%%%%%%%%%%

\section{The case of $2\times 2$ matrices}

The case of $N=2$ is the simplest we can analyze given the structure of our ansatz that is not completely trivial.
The matrices take the form
\begin{equation}
X^+ = %
\begin{pmatrix}
0 & |a_1|\exp(i \omega_1 t) \\
|a_2| \exp(i\omega_2 t) & 0
\end{pmatrix}, %
\quad Z = %
\begin{pmatrix}
z_1 & 0 \\
0 & z_2
\end{pmatrix}
\end{equation}
When we compute the equations that the $z_i$ must satisfy for $a_i,~J$ fixed, we find that
\begin{align}
0 &= z_1 + |a_1|^2(-6 + 4z_1 - 4z_2) + |a_2|^2(-6 + 4z_1 - 4z_2)\\
0 &= z_2 + |a_1|^2(-6 + 4z_2 - 4z_1) + |a_2|^2(-6 + 4z_2 - 4z_1)
\end{align}
Summing the two, we find that $z_1 + z_2=0$, and then we can substitute this result back, to find that
\begin{equation}
z_{1,2} = \pm \frac{6(|a_1|^2 - |a_2|^2)}{1 + 8|a_1|^2 + 8|a_2|^2}
\end{equation}
At this point it becomes clear that the expressions simplify if we consider the two variables $P = |a_1|^2 + |a_2|^2$ and $Q = |a_1|^2 - |a_2|^2$. This is because $P$ is even with respect to the $\BZ_2$ symmetry $|a_1|\leftrightarrow |a_2|$, and $Q$ is odd. Thus, the symmetry algebra acts simply on the variables $P$ and $Q$ themselves.
We find then that
\begin{equation}
\Ecal(J, z_i(|a_i|), |a_i|) = \frac{J^2}{16 P-16 Q}+\frac{J^2}{16 (P+Q)}+\frac{32 (P-1) Q^2}{8 P+1}+2 P
\end{equation}
When we compute the equations that $P,~Q$ must satisfy, obtained by considering $\partial_{P,Q} E=0$, it is convenient to eliminate the $J$ dependence of one algebraic combination of these. After a bit of work, this is accomplished by considering 
\begin{equation}
2 PQ \partial_P E+(P^2+Q^2) \partial_Q E=\frac{4 (4 P-1) Q \left(32 P^3-4 P^2+P \left(32 Q^2-1\right)+16 Q^2\right)}{(8 P+1)^2}\label{eq:factor}
\end{equation}
Notice that this factorizes, so there are three  branches. One where $Q=0$ identically, one where $P=1/4$ identically and another one where 
\begin{equation}
Q^2=\frac{-32 P^3+4 P^2+P}{16 (2 P+1)}\label{eq:Qsol}
\end{equation}
which is positive only if $P\leq 1/4$.

Let us analyze the first one. We can substitute $Q=0$ in $E$,  to find that 
\begin{equation}
E = \frac{J^2}{8P} + 2P
\end{equation}
And the minimum occurs for $P\to J/4$ (here we have taken $J>0$, and obviously $P$ is positive since it is a sum of squares). We can then evaluate that for this solution
\begin{equation}
E = J
\end{equation}
in the limit $J\to 0$, this solution reduces to the trivial solution where all matrices are identically zero. 
For the second solution, we take $P=1/4$, and we find similarly that 
\begin{equation}
E = \frac{J^2 + \left(1-16 Q^2\right)^2}{2 - 32Q^2}
\end{equation}
which is of similar form if we use the variable $x=1-16 Q^2$ (this is, of the form $Ax+B x^{-1}$). Again, the minimum occurs when 
\begin{equation}
Q = \frac{\sqrt{1-J}}{4}
\end{equation}
and we also find that $E = J$ identically. This is the solution where we choose to take $Q>0$. There is a similar solution with $Q<0$ that is a $\BZ_2$ reflection of this solution.
When $J\to 0$, this is the standard fuzzy sphere of $2\times 2$ matrices. When $J\to 1$, this matches our other solution with $Q=0$. The three solutions meet at $J=1$. Beyond $J=1$ this solution does not exist anymore.

In the third branch, we have that
\begin{equation}
E = \frac{2 J^2 (2 P+1)}{(4 P+1) (16 P-1)}-4 P^2+9 P-\frac{9 P}{2 P+1}
\end{equation}
and it is easy to solve for $J$ as a function of $P$ (essentially solving $\partial_PE=0$), giving us
\begin{equation}
J^2=\frac{(1-P) P \left(64 P^2+12 P-1\right)^2}{(2 P+1)^2 (8 P+1)}
\end{equation}
So that we end up with a parametric solution $J(P)$, rather than the other way around. This can be inverted numerically.
This only makes sense if $J^2\geq 0$, so that necessarily $P\leq1$, and remember also that $1/4\geq P\geq 0$ from the reality of $Q$. But moreover, $|Q|\leq P$, and this restricts $P$ to be bigger than $1/16$.
Notice that when $J\to 0$, there are various values of $P$ that can  arise as roots. 
The only one that is new corresponds to $P=1/16$. This is a fuzzy sphere at half radius. This can be easily understood if we make an ansatz of spherical symmetry for a saddle point.
In this case the matrices $X,~Y,~Z$ are all proportional to the corresponding Pauli matrices with proportionality constant $r$. Because the energy for a static configuration is quartic in the matrices, and we have that 
$E=0$ at $r=0$ and $E=0$ at $r=1$, and always $E\geq0$, then the energy must be proportional to $r^2(1-r)^2$. This has a maximum at $r=1/2$, which is the sphere at half radius.
This solution also matches the solution at $Q=0$ that we already had when we set $P=1/4$.
There is similarly a reflected solution with $Q<0$, where we take the other square root branch cut of equation \eqref{eq:Qsol}.
The new solution of the fuzzy sphere at half radius migrates to higher angular momentum as we increase $P$ from $1/16$ to $1/4$ and is a saddle point. Therefore  it has Morse index one. This solution and the one that is reflected by taking $Q\to -Q$ can cancel the two minima from the BPS solution as they meet at $J=1$ with the trivial solution that has $Z=0$ throughout. 

\begin{figure}[ht]
\includegraphics[width=3.0 in]{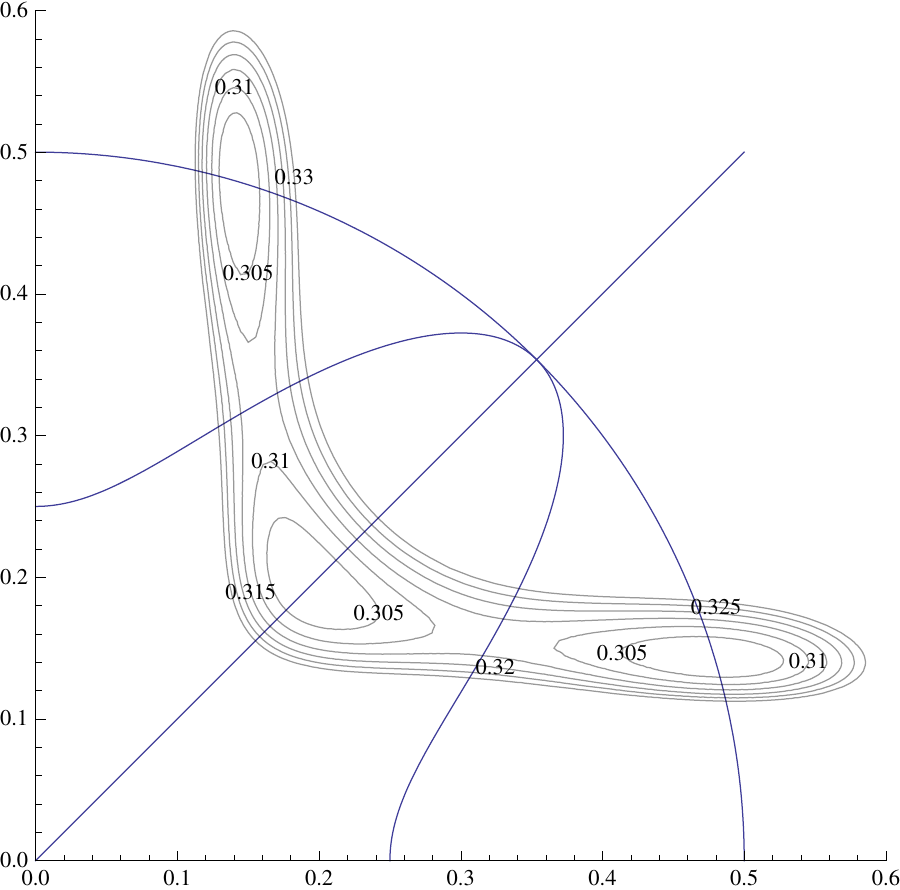}
\caption{Parametric plot of the solutions for $a_1,~a_2$ derived either from equation \eqref{eq:Qsol}, from $P=1/4$ or from the solution $Q=0$ with $a_1=a_2$. 
Superposed we find level sets of the energy function at $J=0.3$, with the energy levels shown, and we see that the curves pass through the critical points of the energy function.}
\label{fig:twotimestwo}
\end{figure}

The full set of solutions is depicted in figure \ref{fig:twotimestwo}. There we can see that for low angular momenta there are five critical points. Three are minima and two are saddles with Morse index one. The saddles and two of the minima are reflected into each other by the $\BZ_2$ symmetry, and one minimum if at the fixed point. The two saddles and two of the minima annihilate each other when $J=1$. Because the minima that annihilate with the saddles are BPS, the saddles need to approach the extremal limit and thus must touch the fixed point set, because one can not descend further from the saddle to the fixed point otherwise.

%%%%%%%%%%%%%%%%%%%%%%%%%%%%%%%%%%%%%%%%%%%%%%%%%%%%%%%%%%%%%%%%%%%%%%%%%%%%%%%

\section{The case of $3\times 3$ matrices}

The equations that the $a_i$ and $z_i$ need to satisfy can be directly derived from \eqref{eq:momentumdot} and \eqref{eq:nn}. We have not been able to solve them algebraically in general, although there is one trivial solution with $z_i=0$ and all $a_i$ equal to each other. This solution exists for any value of the angular momentum (and actually all values of $N$ as well).
Our strategy for solving the problem is to start with known solutions at $J=0$ and perturb them numerically slowly by varying $J$ until a new solution near the old one is found. Because the solutions are saddle points of the energy function, which is considered as a Morse function, small perturbations of the function preserve the saddles in general.  
These are saddles for any $J$ until a subset of the saddles collides. A saddle with index $m$ can be annihilated by a saddle of index $m+1$ or $m-1$
\footnote{Recall that the index of a critical point is number of negative eigenvalues of the Hessian at that point.}. 

\begin{figure}[ht]
\includegraphics[width=3.0 in]{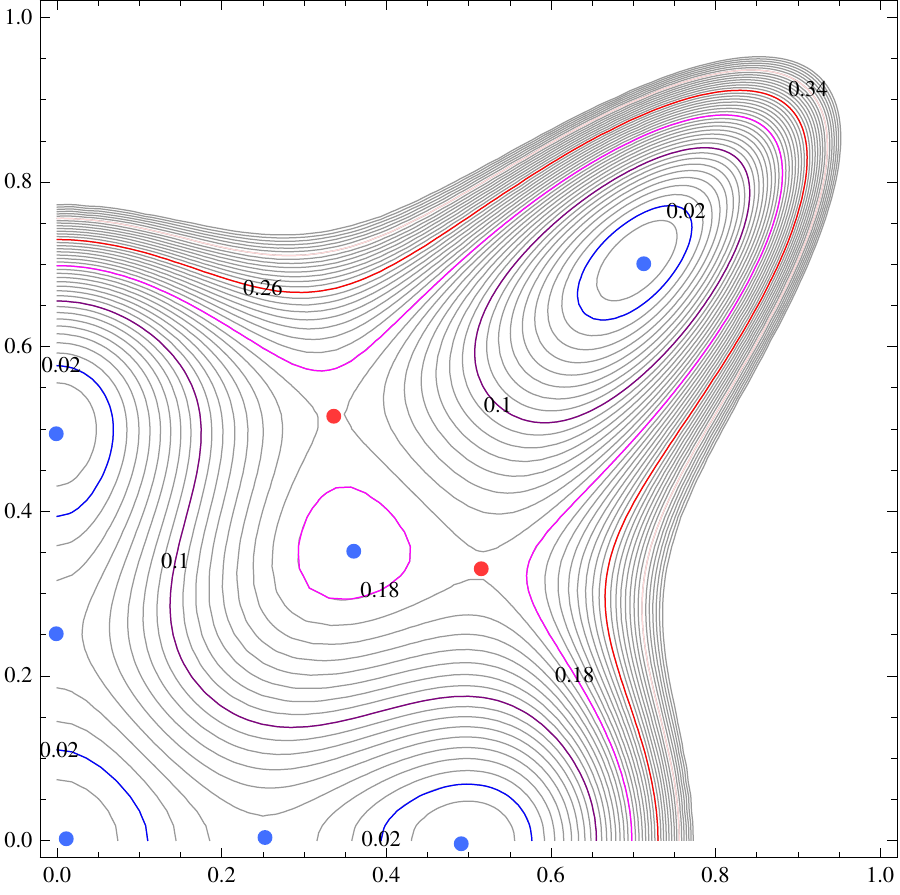}
\caption{Energy function at $J=0,~|a_3|=0$. The axis indicate $|a_1|,~|a_2|$, and the colored contours have their energy values indicated. The solid dots indicate approximate positions for the various saddles. The solid dots in red indicate new saddles that do not arise from collections of round fuzzy spheres.}
\label{fig:energy3}
\end{figure}

There are obvious solutions at $J=0$. These are the fuzzy sphere vacua. For any such fuzzy sphere with $k\times k$ matrices, we can also find an unstable fuzzy sphere at half size, which is analogous to the one we found for $2\times 2$ matrices. We can then combine these solutions together into new solutions. The reason for this is that in any fuzzy sphere for $k\times k$ matrices we have $|a_k|^2 = 0$ and the matrix is upper triangular. This is true regardless of if the fuzzy sphere is stable or unstable.  We can then mix and match these solutions and put them in some order. For $3\times 3$ matrices this does not matter as there are not too many ways of partitioning $3$ into integers, but the strategy works in general for other values of $N$. 
The value $J=0$ is technically a singularity of the family of morse functions, because the fuzzy sphere vacua end up with some $a_i=0$, and we argued before that these points need to be identified with each other in the one point compactification of the open intervals $|a_i|\in (0,\infty)$ to get a sphere topology. Numerically, we can start with these solutions and perturb entries that start at zero by a small amount, while at the same time turning on a small amount of angular momentum. All solutions persist under this procedure, so we can safely describe them by taking the limit $J\to 0$. We also need to count them with multiplicity, because we have our dihedral group of (quantum) symmetries that let us find new solutions of the $a_i$ by permuting them clockwise, or by reflection. This is described in table \ref{tab:fuzzysp}
\begin{table}[ht]
\begin{tabular}{|l|l|c|c|c|}
\hline Splitting & Stability & Morse Index & Multiplicity & Unbroken Quantum Symmetry\\
\hline 1+1+1 & S, S, S & 0&1&$D_3$\\
 2+1 & S,S & 0 & 3&$\BZ_2$\\
 2+1 & U, S & 1 & 3&$\BZ_2$\\
 3 & S& 0 & 3&$\BZ_2$\\
 3 & U & 2 &3&$\BZ_2$\\
 \hline
\end{tabular}
\caption{Table of sphere solutions. The splitting indicates the size of matrices of the fuzzy spheres, $S,U$ indicates if they are stable BMN vacua, or unstable spheres at half size, the Morse index is the number of negative modes of the Hessian (after perturbing by a small $j$, and the multiplicity is the number of copies of the solution that are obtained from using the group symmetry actions on a  given solution.}
\label{tab:fuzzysp}
\end{table}
If we take the solutions as above and we compute the Morse polynomial with just these solutions, we find that
\begin{equation}
M_{trial}(t)= 1+3+3t+3+3t^2+t^3=7+3t+3t^2+t^3
\end{equation}
where the last entry (the one for $t^3$) corresponds to the maximum of the energy function at infinity.
The Poincare series of the three sphere is
\begin{equation}
P(t) = 1 + t^3
\end{equation}
The Morse inequalities require that $M_{trial}(t) - P(t) = (1+t)Q(t)$ where $Q$ should be a polynomial with positive integer coefficients. In particular we should have that $M(-1) = P(-1)$. This is not the case. This indicates that we are missing critical points of the energy function. Since all the solutions above have an enhanced $U(1)$ rotation symmetry, it makes sense to look for solutions with such a rotation symmetry for more saddles. We do this graphically in figure \ref{fig:energy3}. In the figure the $z_i$ have already been solved for, but the $|a_i|$ are variables.

We see from the figure that there are additional saddles that are not reflection symmetric with respect to the diagonal, nor are they on the edges of the graph. 
These new saddles have index $1$, and the orbit under the symmetry group produces $6$ saddles in total, and no unbroken symmetry. These would add an additional $6t$ to $M_{trial}(t)$. With this additional set of solutions we find that
\begin{equation}
M(t)= 1 + t^3 + (1+t)(6+3t)
\end{equation}
and now we satisfy the Morse inequalities. This means that it is consistent if these are all the saddle points and we are not missing any.
 
We have not found any other saddle point numerically at $J=0$. If they exist they should have all three $|a_i|\neq 0$. It is possible to use this information to show that no such saddle can exist, and not just for $N=3$ but for all $N$.
 
The energy function is given by
\begin{align}
\Ecal(J,|a_i|,z_i) &= \frac{J^2}{8N^2}\sum_{i=1}^N \frac{1}{|a_i|^2} + \sum_{i=1}^N \frac{1}{2}(z_i + 2|a_{i-1}|^2 - 2|a_i|^2)^2 + 2(1 + z_{i+1} - z_{i})^2|a_i|^2
\end{align}
The equations of motion for the $a_i$ yield
\begin{equation}
\label{eq:eoma}
0 = -\frac{J^2}{4N^2 |a_i|^3} + 4|a_i|\left( (1 + z_{i+1} - z_{i})^2 + (z_{i+1} - z_{i}) + 2(-|a_{i-1}|^2 + 2|a_{i}|^2 - |a_{i+1}|^2)\right)
\end{equation}
Next we suppose that $|a_i|\neq 0$ for all $i$ at $J=0$.
The first term of \eqref{eq:eoma} vanishes and we may divide the rest by $4|a_i|$.
We are left with
\begin{equation}
0 = (1 + z_{i+1} - z_{i})^2 + (z_{i+1} - z_{i}) + 2(-|a_{i-1}|^2 + 2|a_{i}|^2 - |a_{i+1}|^2)
\end{equation}
Summing over all $i$ we have
\begin{equation}
0 = N + \sum_{i=1}^N (z_{i+1} - z_{i})^2
\end{equation}
Since the summation is non-negative and $N$ is positive, we have reached a contradiction.
Thus $|a_i| = 0$ for at least one $i$ at $J = 0$.

The equation of motion for the $z_i$ yields.
\begin{equation}
\label{eq:eomz}
0 = z_i - 4|a_{i}|^2\left(\frac{3}{2} + z_{i+1} - z_i\right) + 4|a_{i-1}|^2\left(\frac{3}{2} + z_{i} - z_{i-1}\right) 
\end{equation}
Summing these equations yields a traceless condition, $\sum_{i} z_i = 0$.
This had to be true because $\dot{P}_Z = 0$ and the trace of the matrix model is just a harmonic oscillator.

This feature has the following implication. When we perturb away from $J=0$, the $|a_j|^2$ will be modified very slightly (at order $J^2$), but for the one that begins at zero $|a_{\ell}|^2$ it is different.
In the kinetic term we will have a singular term, so that the energy function will look like
\begin{equation}
\Ecal(J)\simeq \Ecal(0) + \alpha |a_\ell|^2 + \frac{J^2}{8N^2 |a_\ell|^2}
\end{equation}
and clearly the $|a_{\ell}|$ that minimize this are such that $|a_\ell|^2 \simeq J/\sqrt{8 N^2 \alpha} $. When we plug this into the energy function we find that there is always a linear term in $J$. This means that the fuzzy configuration built with out ansatz can never be considered as a rigid body; for rigid bodies the energy goes like $J^2$, where the coefficient of proportionality depends on the moment of inertia. The fuzzy configurations rotate by turning on wave-like excitations on the sphere. This is exactly as expected for a membrane.
 
\begin{figure}[ht]
\includegraphics[width=3.5 in]{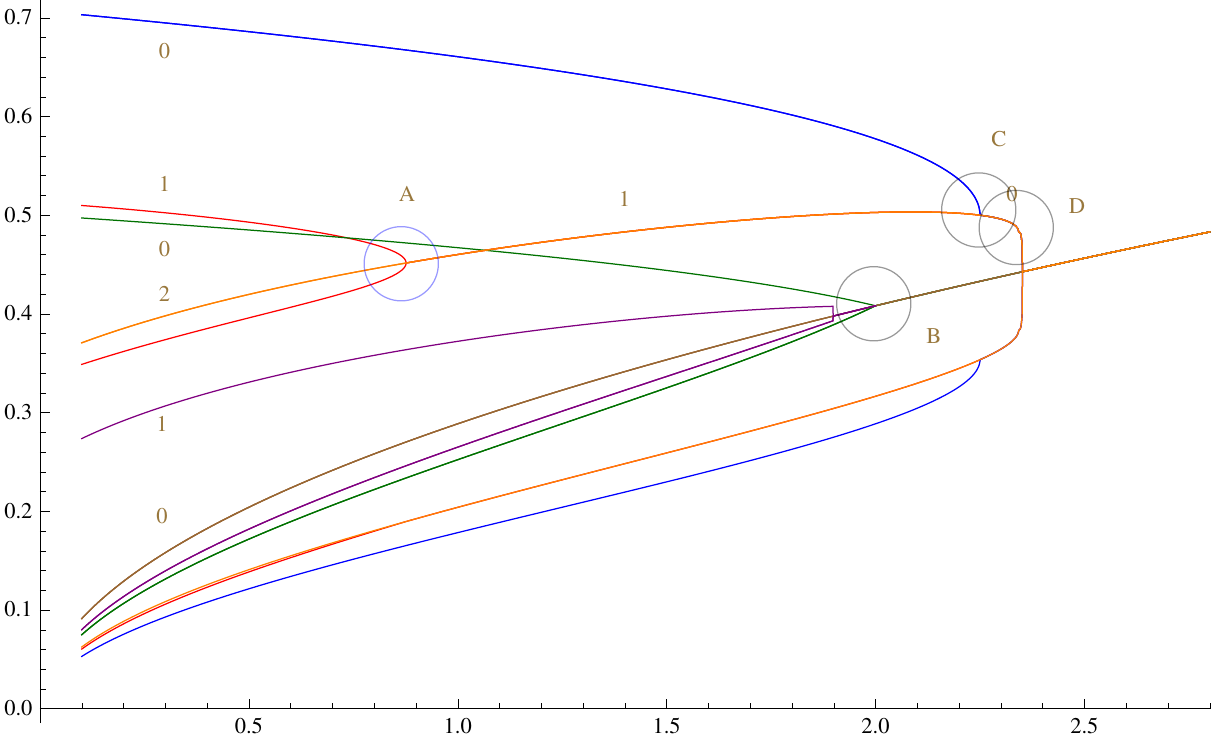}
\caption{Phase diagram of solutions as a function of $J$ on the abscissa, and the $|a_i|$ are plotted on the ordinate. We follow the solutions slightly perturbed from $J=0$. Shown are the values of the $|a_i|$. Different solutions for $|a_1|,~|a_2|,~|a_3|$ are colored differently, but all values are shown in the same color for the same solution. The numbers attached to saddles are the Morse index. Phase transitions where solutions merge or end are shown as $A,~B,~C,~D$. The ones that are not marked below the maximal symmetry solution are $C,~D$ also.}
\label{fig:phdiag1}
\end{figure}

\begin{figure}[t]
\includegraphics[width=3.0 in]{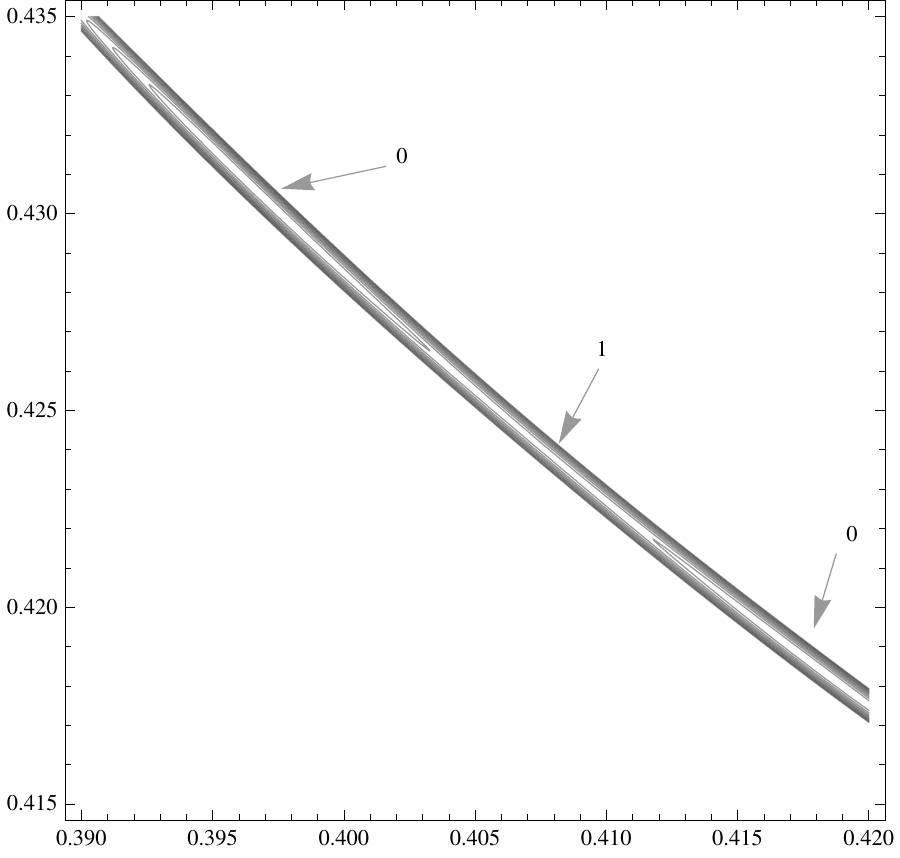}
\caption{Energy contours in $|a_1|$, $|a_2|=|a_3|$ at $J=2.1$. The saddles are hard to see. The energy contours have energies starting at $E=2.1$ and spaced at $\delta E = 2\times 10^{-7}$ showing ten contours.} \label{fig:saddle_search}
\end{figure}

A presentation of the solutions found at $J=0$ for $3\times 3$ matrices, followed as we change $J$ is shown in figure \ref{fig:phdiag1}. 
As shown in the figure, there are many phase transitions. The one marked $A$ corresponds to two saddles of index one and one saddle of index 2 merging into a single saddle of index 1. This is repeated in three different locations due to the symmetry operations. The unbroken symmetry of the incoming saddles of index one changes from all symmetry broken to a saddle with a $\BZ_2$ symmetry unbroken. It also shows that the saddle of index one ends smoothly and without corners. Actually this is what is expected in general: saddles should end smoothly and without corners as we pass through critical values of $J$. The phase transition marked as $C$ is inconsistent as shown: a single saddle of index zero and a saddle of index one can not annihilate into a saddle of index zero (this would violate the Morse inequalities). This means that there is a solution missing. Similarly, in phase transition $D$, the line actually ends. The vertical line down is an artifact of joining the numerical solutions, and instabilities of the numerical method when the Hessian is degenerate. This solution ending also indicates a missing solution, because of inconsistency with the Morse inequalities. Finally, at the transition $B$ we see that a set of three saddles of index zero and index one merge with the saddle with all symmetries unbroken. Although this is in principle allowed, it is not smoothed out like in transition $A$, or as one would expect from transition $C$ once a solution is found. We can conjecture that the solutions that correspond to the green and purple lines in the figure actually continue to the other side. Because of the way solutions are merging and the dihedral symmetry being restored, transition $B$ is a multi critical point. It is natural to believe that the solutions that are continued end on the other inconsistent transitions. They should have also an unbroken $\BZ_2$ symmetry each and come in three copies. Because of the unbroken $\BZ_2$ symmetry, two of the $|a_i|$ should be equal to each other. This makes it possible to guess the solutions by looking at energy contours in a two dimensional plot where we set $|a_3|=|a_2|$ a bit after the transition at $J\simeq 2.1$. We should be looking for a $BD$ line of index one (it's the only way we can cancel a solution of index zero). The green line actually saturates the BPS inequality, and this should persist in the analytic continuation to the right. This indicates that there should be a saddle of index 0 joining $BC$. This process is depicted in figure \ref{fig:saddle_search}
\begin{figure}[t]
\includegraphics[width=3.5 in]{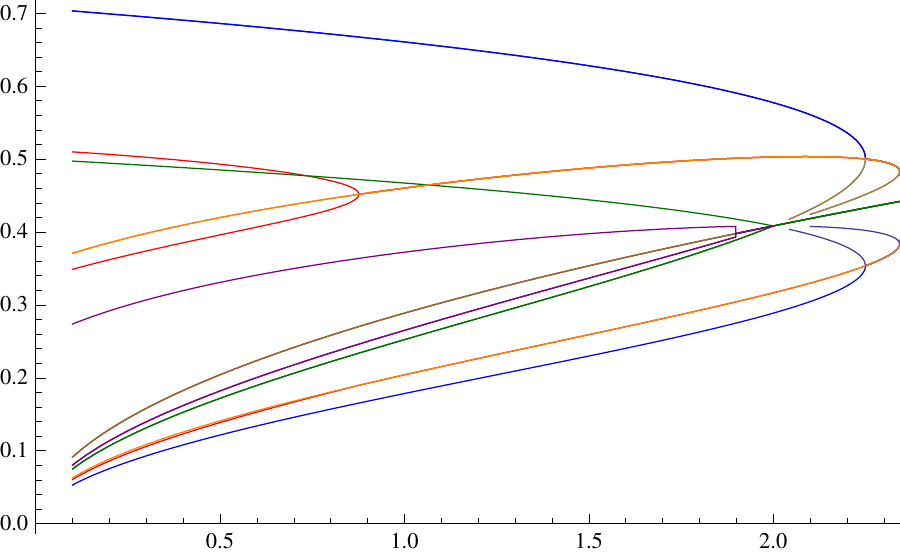}
\caption{Full Phase diagram of solutions as a function of $J$ on the abscissa, and the $|a_i|$ of a single solution are plotted on the ordinate.}
\label{fig:phdiag2}
\end{figure}
Once we include the new saddles after this point, we can complete the phase diagram in figure \ref{fig:phdiag2}.
What we see from the figure is that first, there is a maximal angular momentum after which there is only one solution, and this solution preserves the maximal dihedral symmetry. Second, the trajectory of the maximal sphere goes through a phase transition at a finite $J$ before reaching this maximal angular momentum. We have checked that after the first transition the solution stops being extremal (BPS), even though it is a local minimum. This means that the corresponding phase can be considered metastable from energetic considerations. 

At the maximal angular momentum, it becomes classically unstable. The jump from the metastable to the stable configuration is a first order transition, but at the place where it changes from being BPS to being non-BPS, we have coexistence of a second order phase transitions and a first order phase transition with the same energy. This degeneracy is expected to be lifted by quantum corrections. At small $\hbar$ we expect that the phase diagram has the same structure we have shown, because it is controlled by topological aspects of a Morse function.
This is similar to the transition found in \cite{Armas:2012bk}, where studying spherical D-brane configurations, beyond the maximal giant graviton there is a non-BPS metastable D-brane solution that continues for a while with larger R-charge (this branch was first found in 
\cite{Grisaru:2000zn}, but the stability analysis was not done), and the family of solutions ends in a transition that should take us to another branch. This transition should be where the metastability is lost.

%%%%%%%%%%%%%%%%%%%%%%%%%%%%%%%%%%%%%%%%%%%%%%%%%%%%%%%%%%%%%%%%%%%%%%%%%%%%%%%

\section{Other Examples}

Our next task is to understand how to go beyond $N=3$. Armed with the information that at least one of the $|a_i|$ needs to be zero at $J=0$, suggests that we can begin by looking at saddles for smaller $N$ and fit them into saddles for $N$ by either bordering by zeros, or by taking direct sums of solutions for smaller $N$ such that the rank adds up to $N$. We choose these saddles to have $|a_k|=0$ for the entry on the first row at the bottom corner.
These will all fit the ansatz where $X^+$ is upper triangular, bordering the diagonal, and with a zero in the bottom corner. This actually always produces correct solutions at $J=0$.
From equation \eqref{eq:eomz}, we find that for the values of $|a_i|$ such that $|a_i|=0$, $z_i$ only depends on the previous ones, and therefore decouples from $z_{i+1}$, and similarly $z_{i+1}$ will be related to $z_{i+2}$, but not the other ones. Secondly, the equations for $|a_i|$ are trivially satisfied in this case, because of the $|a_i|$ appearing in front of it in equation \eqref{eq:eoma}. The rest of the equations are satisfied if they were satisfied for smaller values of $N$. This gives us an ample trove of solutions with which we can explore the phase diagram.
\begin{figure}[t]
\vspace{-1.8cm}
\includegraphics[width=2.5 in]{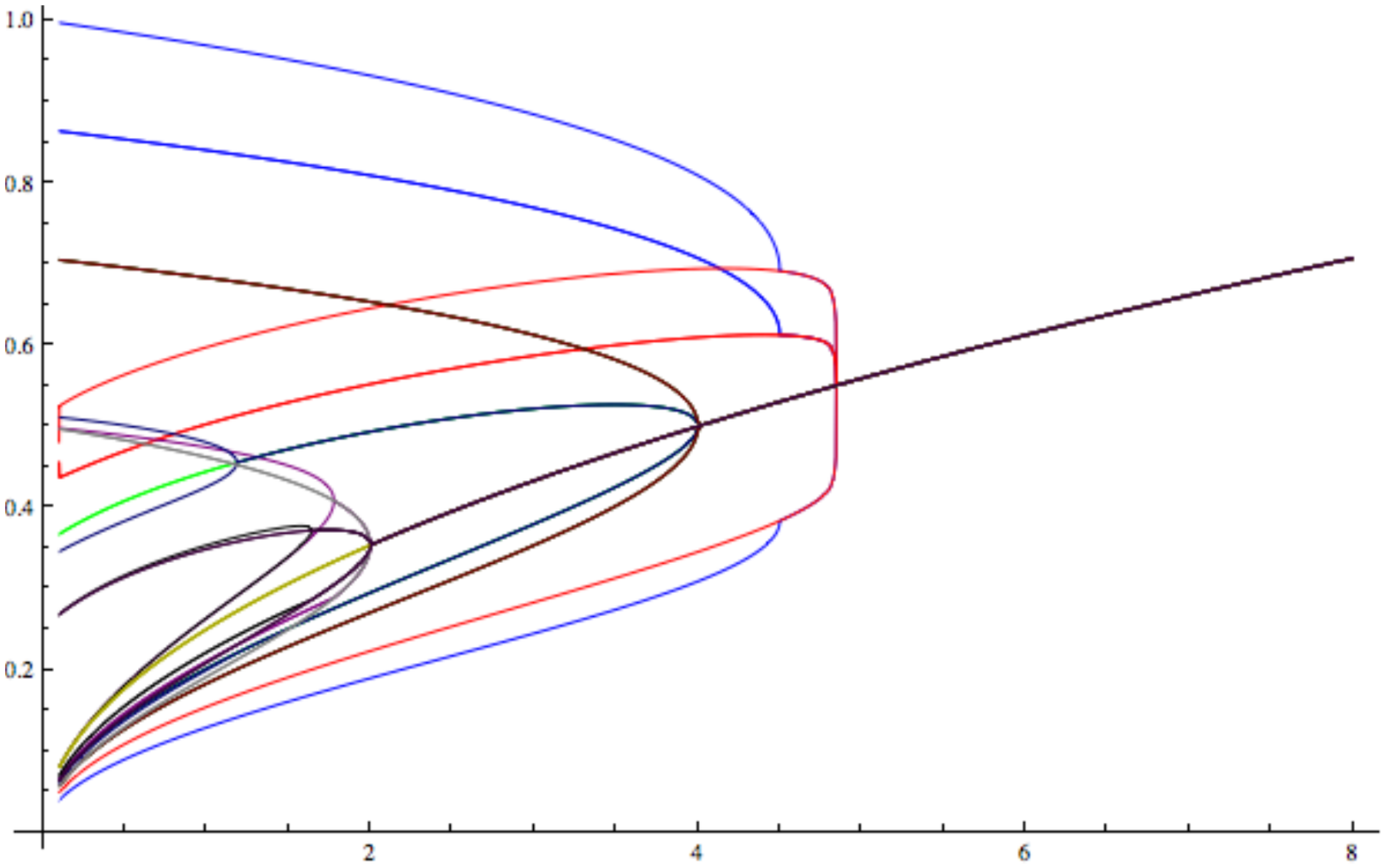}%
\quad\includegraphics[width=2.5 in]{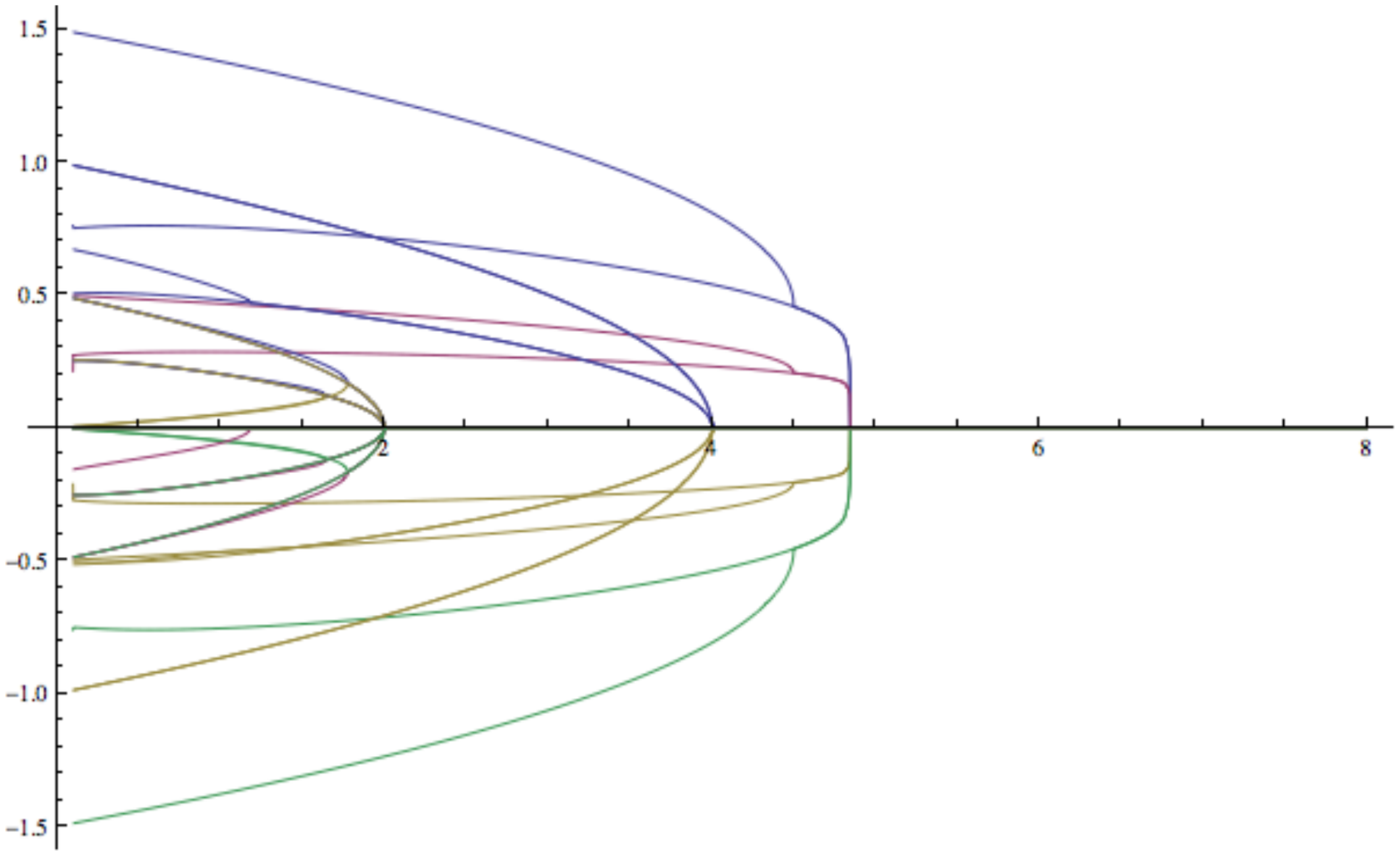}
\vspace{-2cm}
\caption{On the left, partial phase diagram of $4\times4$ solutions as a function of $J$ on the abscissa, and the $|a_i|$ of a single solution plotted on the ordinate in the same color. Notice the similarity of the transitions to the ones for $3\times 3$ matrices in figure \ref{fig:phdiag1}. On the right, we evaluate also the $z_i$ values for the given solution of the $|a_i|$. The figure is not symmetric under $z_i\to -z_i$ which indicates that many of the phases are not `parity invariant' with respect to the dihedral discrete symmetry group.}
\label{fig:4x4}
\end{figure}
This is explicitly shown in figure \ref{fig:4x4}.
We see qualitatively the same structure. There is a maximal angular momentum beyond which only the trivial solution exists, and it ends in a first order phase transition with a metastable phase. Because there are many more solutions at $J=0$ at each $N$ (the number is bounded below at least by twice the partitions of $N$), the full phase diagram is more complicated as we increase $N$, and it is clear from the Morse analysis that we are missing quite a number of saddles.
Although two new solutions always exists for any $N$, the maximal fuzzy sphere and the sphere at half size, one can expect in general that there are quite a number of intrinsically new (indecomposable) solutions that appear at any $N$.

\begin{figure}[b]
\includegraphics[width=3.5 in]{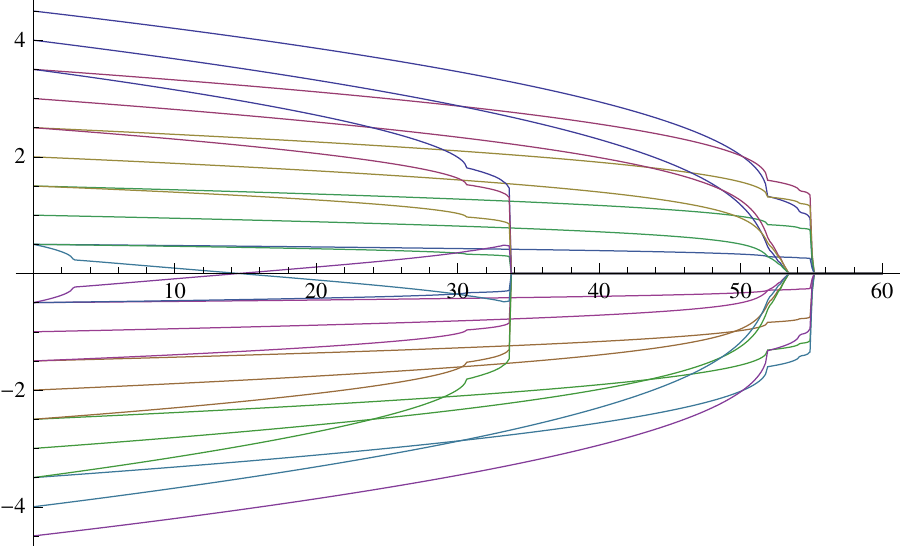}
\caption{We evaluate $z_i(J)$ for three different solutions of $10\times 10$ matrices, the maximal fuzzy sphere, a near maximal fuzzy sphere bordered by zero, and a solution for a  fuzzy sphere of spin $7/2 \oplus 1/2$.}
\label{fig:fuzzy10}
\end{figure}

For illustration purposes, we also show some of the solutions for $10\times 10$ matrices in figure \ref{fig:fuzzy10}. This shows the procedure for building up solutions by bordering by zero or adding up previous solutions in more detail. It is clear that the full phase diagram is quite complicated, some of the solutions ending in first order phase transitions, and some others ending in what appear to be multi-critical points. Here we basically show that the problem is amenable to computer calculations.
We should point out that for larger $N$ we do not have a good strategy to search for the intrinsically new solutions yet.
The graphical method that worked for $N=3$ is unsuited for higher dimensions, as we can not visualize the data.

%%%%%%%%%%%%%%%%%%%%%%%%%%%%%%%%%%%%%%%%%%%%%%%%%%%%%%%%%%%%%%%%%%%%%%%%%%%%%%%

\section{Large $N$}
\label{sec:largeN}

An interesting way to proceed to large $N$ is the following. Since the energy function in equation \eqref{eq:nn} is of nearest neighbor type, we can think of it as a discretized version of an energy which is an integral of an energy density. Indeed, this is what the interpretation of the solutions as of matrix mechanics as discretized membranes indicates we should be doing \cite{Hoppe,de Wit:1988ig}. With that in mind, we want to replace $z_i\to  \tilde z(\theta)$ and the same for $|a_i|^2\to |\tilde a|^2(\theta)$, where $\theta$ is a periodic coordinate with period one, rather than a discrete set with $N$ elements. We basically take $i \simeq N \theta$. Then expressions of nearest neighbor differences get replaced by derivatives $z_{i+1}-z_{i}\to N^{-1} \partial_\theta \tilde z$. We also need to remember that the maximal  fuzzy spheres are  of size $N$ when we have $N$ D0-branes, and we want to rescale this out of the energy. Therefore, we write $z_i\simeq N z(\theta)$, and $|a_i|^2 \simeq N^2 |a|^2(\theta)$ and $\sum_i= N\int d\theta$. When we do this, we find that
\begin{equation}
V_{pot} = V(|a_i|,z_i) \to N^3 \int d\theta\left[\frac{1}{2}  \left(z - 2\partial_\theta |a|^2 \right)^2 + 2\left(1+\partial_\theta z\right)^2|a|^2\right]
\end{equation}
and similarly, the  terms that contain angular momentum, when we rescale $J\to N^3 \tilde{\jmath}$, give us
\begin{equation}
E_{kin} = N^3\int d\theta \frac{\tilde{\jmath}^2}{8|a(\theta)|^2}
\end{equation}
The point is that in this rescaling, we get a common factor of $N^3$ in front of the energy that can be dropped if we want to, to obtain a classical membrane energy which is independent of $N$. 
The value of  $N$ can be changed effectively by changing the periodicity of $\theta$ without changing the energy density further. We use the relation $N_{tot} = N \int d\theta$ to convert the changing period of $\theta$ into a different value of $N$, and we can always change scales by taking $\theta\to \alpha\theta$, $z\to \alpha z$ and $|a|^2\to \alpha^2 |a|^2$.
The energy function also has a $\BZ_2$ symmetry where we change $z\to -z$ and $\theta\to -\theta$.

The energy function is now a local integral of the functions $z,a$ and their derivatives.
The condition to be a critical point of the effective energy is a pair of differential equations, one for $z$ and one for $a$ that come from the variational principle.
A complete set of initial conditions requires specifying $z(0),~a^2(0),~\partial_\theta z(0),~\partial_\theta a^2(0)$. These can be evolved in $\theta$, but the trajectories need to be periodic with a fixed prescribed period (which we are choosing to be set equal to one). This effectively quantizes the set of possible solutions so that they are discrete. 

Notice that the system is translation invariant in $\theta$, thus there is a trivial one parameter family of solutions that is obtained by translation. This becomes a  $U(1)$ symmetry that was only realized as a $\BZ_N$ quantum symmetry at finite $N$, and which is spontaneously broken on most of the solutions. Because $|a^2|$ is bounded for interesting solutions, we can always start out from a place where $\partial_\theta |a|^2=0$.

The fuzzy spheres at zero energy appear at $\tilde\jmath=0$, and are at the zeros of $V_{pot}$. These occur at
\begin{align}
\partial_\theta z &= -1 \\
\partial_\theta |a|^2 & = \frac{1}{2} z
\end{align}
so that they satisfy a  first order set of equations (typical of BPS states), rather than the usual second order equations.
At finite $N$, the $Z$ matrix for fuzzy spheres is the matrix of spins in an $N$ dimensional representation of $\mathfrak{su}(2)$. Consecutive matrix elements differ by $1$ and so the equation for $z(\theta)$ is capturing this effect.
If we start from $z=0$ at $\theta=0$ (so that $z(\theta)=-\theta$), and  $|a|^2(0) = a^2_{max}$, it is straightforward to integrate these and find that
\begin{equation}
\partial_\theta |a|^2= -\frac{1}{2} \theta
\end{equation}
so that 
\begin{equation}
|a|^2(\theta)= a^2_{max} - \theta^2/4
\end{equation}
we hit $a^2(\theta_0)=0$ at some finite value of $\theta_0$. This is a singular point in the differential equation system because the effective derivative term squared  for $z$ vanishes at this point. We 
are allowed to have large jumps in $z$ at this point because there is no energy cost to it. We recover this way that the set of possible ground states is a collection of fuzzy spheres.
Quantization then requires that each of these have an integer amount of D0 brane charge. This is reminiscent of the LLM droplet picture \cite{Lin:2004nb}, where quantization of the area arises from a Dirac quantization condition.
Notice that the solution can also be written as 
\begin{equation}
|a|^2(\theta)= a^2_{max}-z^2(\theta)
\end{equation}
which gives $|a|^2+|z^2|= \text{const}$, as one expects from a sphere written in cylindrical coordinates (the additional angle is associated to the $\BZ_N\to S^1$ rotational invariance of the set of solutions we are considering in the large $N$ limit).

At this point, going beyond the BPS solutions, we want to change perspective and think of the variable $\theta$ as a time coordinate, and the effective energy function we had before as a Lagrangian whose variational principle gives some non-trivial equations of motion. Using this change of point of view, we see that the Lagrangian has a (repulsive towards infinity and the origin of $|a|^2$ when $\tilde j \neq 0$) potential of the form
\begin{equation}
V_{eff} \simeq -\frac{1}{2}|z|^2 - 2|a|^2 - \frac{\tilde{\jmath}^2}{8|a|^2}
\end{equation}
there is a non-trivial curved metric
\begin{equation}
ds^2 = 4 (d|a|^2)^2 + 4|a|^2 dz^2
\end{equation}
with translation symmetry in $z$
and a non-trivial magnetic potential associated to the one form
\begin{equation}
\Acal = -2z d|a|^2 + 4 |a|^2 dz
\end{equation}
which in these coordinates produces a constant magnetic field. Notice however, that because the metric is curved, the magnetic field per unit normalized area actually changes and becomes weak when $|a|^2$ gets large. The magnetic field will try to bend trajectories into confining circular orbits, but it has to compete with a repulsive potential that tries to destabilize the system. Also, the metric has a scaling symmetry where $|a|^2\to \gamma^2 |a|^2,~z\to \gamma z$. The metric is also positively curved away from $|a|^2=0$, which is a singularity. If we thought of the $z$ as an angle coordinate and the $|a|^2$ as a radial variable, then the curvature wants to repel geodesics from hitting $|a|^2=0$. 

The obvious critical point of the equations of motion where nothing moves, is at the maximum of the potential and is in unstable equilibrium. This produces a periodic orbit for any period. Other periodic orbits need to be found by trial and error, and once a sufficiently approximate solution of the periodicity condition is found it is possible to zoom into it.

\begin{figure}[t]
\includegraphics[width=2.9 in]{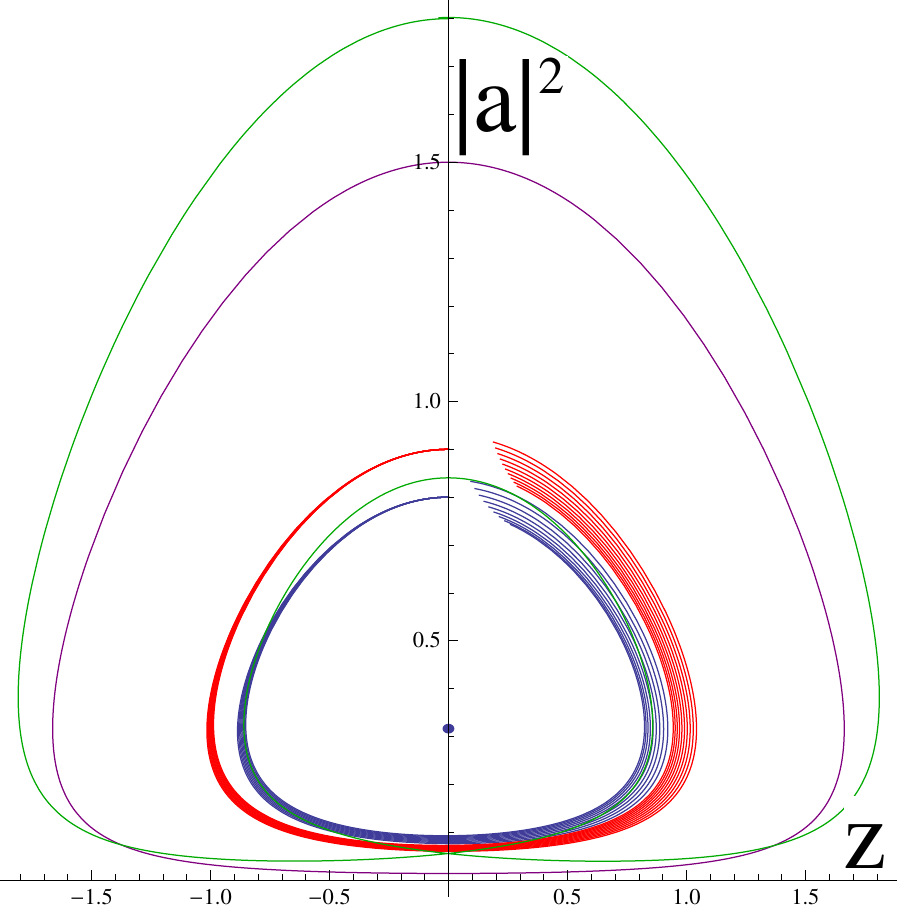}
\caption{Parametric plots of $(z(\theta),~|a|^2(\theta))$ for various initial conditions at fixed value of $\tilde{\jmath}^2 = 0.1$. We show a BPS trajectory in purple, and examples of scanning over parameters to find periodic trajectories  in blue and red. The fixed point is marked. The green solution which was found by scanning winds twice around the fixed point and it is not BPS.}
\label{fig:largeNt}
\end{figure}

Particularly simple examples of this search can be performed if the solutions are reflection symmetric under $z\to -z$. Then the solution can be characterized by the value of $|a|^2(\theta_0)$, when $z(\theta_0)=0$, and the condition for symmetry forces $\partial_\theta |a^2|(\theta_0)=0$. We can then move the velocity of $\partial_\theta z(\theta_0)= \xi$ as a scanning parameter.  Some examples of this procedure are depicted in figure \ref{fig:largeNt}.

When the orbit returns to $z(\theta_1)=0$ at some later time (not necessarily the first time around), we can the compute $\kappa=\partial_\theta|a|^2(\theta_1)$
from solving the equations of motion, and as we scan over $\xi$ we look for sign changes in $\kappa$. We can then zoom in for the parameter $\xi$ that solves the periodicity condition. The point of intersection can be 
the same one as before or it can be different.
If it is different we double the time of this half orbit, and it becomes periodic.

It is also interesting to study the first order differential equations that arise for states that saturate the BPS inequality at finite $J$. These equations are given by
\begin{align}
\partial_\theta z &= -1 + \frac{\tilde\jmath}{4|a|^2} \\
\partial_\theta |a|^2 &= \frac{1}{2} z
\end{align}
and are obtained from requiring the vanishing of the squares in equation \eqref{eq:Hsq} after substituting our ansatz. 
These can in turn be derived from a variational principle for an auxiliary Lagrangian of the form
\begin{equation}
L = \dot{q}^2 - q + \frac{\tilde{\jmath}}{4}\log q
\end{equation}
where $q=|a|^2$, and $z$ plays the role of the canonical conjugate of $q$ (namely $z= 2 \dot q=p_q$). The effective one dimensional potential ${\cal U}(q)= q - \frac{\tilde{\jmath}}{4}\log q$ is bounded from below and goes to infinity at $q\to0$ and also at $q\to \infty$, so it produces automatically closed periodic orbits without self intersections. There is also only one minimum at $q = \tilde{\jmath}/4$.

These trajectories are interpreted geometrically as multiply wrapped tori that wrap the same torus, where the wrapping number is the number of times we have to go around the orbit so that the period matches the number of D0-branes. 

The period of the orbit can be calculated using standard techniques as follows:
\begin{equation}
\Theta = \oint \frac{dq}{\dot q}\propto \oint \frac{d q}{z}
\end{equation}
where the integral is over a periodic orbit, which is characterized by the parametric equation
\begin{equation}
\label{eq:curve}
{\cal W} = \dot{q}^2 + q - \frac{\tilde{\jmath}}{4}\log(q)= z^2 + q - \frac{\tilde{\jmath}}{4} \log(q)
\end{equation} 
where ${\cal W}$ is a constant of integration (the energy associated to the lagrangian function $L$). This solves for $z$ as a function of $q$ readily.
Indeed, the parameter ${\cal W}$ this way defines a  curve with a differential, and then $\Theta$ is the period integral over the differential. We have to be careful with this  interpretation: the curve is real analytic and not a complex curve. More general solutions for the BPS states in the continuum limit were found in \cite{Bak:2005ef}, and they similarly show up with various logarithms \footnote{The variable $W$ in their work is related to $q$ in ours, with $q\simeq W^2$}.

\begin{figure}[t]
\includegraphics[width=3.5 in]{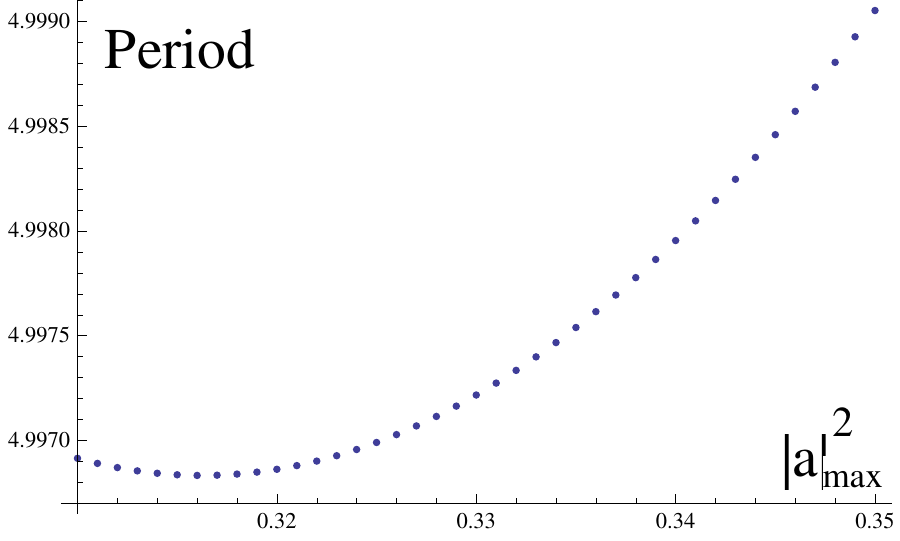}
\caption{Period of an orbit as a function of $|a|^2_{max}$ at $z=0$, and at fixed $\tilde \jmath^2=0.1$. }
\label{fig:period}
\end{figure}

Because ${\cal U}(q)$ has a non-trivial third derivative about the minimum, it is possible to show that the period for orbits very near the fixed point have a decreasing period as we go away from the fixed point, and for large orbits, the period increases again. This is depicted in figure \ref{fig:period}.

Since we need to fix the period to fix the number of D0-branes, while we are allowed to vary $\tilde \jmath$, this shows that for some values of $\tilde \jmath$ there is more than one non-trivial solution (the trivial  fixed point orbit can have any period we want to). The larger solution is interpreted as a fat torus which in principle can become very large as we decrease $\tilde \jmath$ going smoothly to the fuzzy spheres in the limit $\tilde\jmath\to 0$ 
 while the small one is a thin torus that gets thinner and disappears into the trivial solution at finite $\tilde \jmath$. 

As we increase $\tilde \jmath$, we increase the period of the orbits near the fixed point, as well as the minimal period. Eventually the period is too long and the BPS solutions with the fixed period we want disappear. This behavior can already be seen for  the phase diagram of $3\times 3$ matrices in figure \ref{fig:phdiag2}, where one of the new family of minima joining transitions $B$, $C$ in figure \ref{fig:phdiag1} plays the role of the small torus solution whereas the maximal fuzzy sphere family  plays the role of the large torus. This phenomenon can happen for any multiply wound torus in the same way. At the transition $C$ two BPS  minima end together by joining with a saddle and becoming a non-BPS minimum. Hence, it would not be captured by the BPS solutions we are finding.

%%%%%%%%%%%%%%%%%%%%%%%%%%%%%%%%%%%%%%%%%%%%%%%%%%%%%%%%%%%%%%%%%%%%%%%%%%%%%%%

\section{Topology change}

Now that we have the solutions for the matrix ansatz, even if computed numerically, we can analyze the geometry of the resulting matrix configurations as membranes using the ideas developed in \cite{Berenstein:2012ts}. The main idea in that work is that when we add a probe brane (more precisely, we take a direct sum of the matrix configuration and a configuration of $1\times 1$ matrices) and we then consider fermionic off-diagonal modes connecting the probe to the matrix configuration.
The resulting Hamiltonian is a truncation of the BFSS matrix model to a version with less supersymmetry \cite{Banks:1996vh} from which we can obtain a non trivial surface embedded in $\BR^3$ for any set of three Hermitian matrices. The fermions end up carrying a non-trivial topological structure that forbids a global splitting into positive and negative modes everywhere and thus there have to be degenerations. The geometric locus where this is not possible is special in a physical sense; the fermions can not be integrated out. When we cross the special locus the number of positive eigenvalues and negative eigenvalues of an effective Hamiltonian changes. This spectral flow of eigenvalues past zero defines an integer valued index that colors $\BR^3$. The interfaces where the index changes defines the membrane. This is related to the Hanany-Witten effect \cite{Hanany:1996ie}, and more precisely, the creation of fundamental strings by crossing D-branes \cite{Danielsson:1997wq}.

The main calculational tool is to look at the spectrum of the following Hermitian operator which depends on the position of the probe characterized by the vector $\vec{\xi}\in\BR^3$
\begin{equation}
\hat {\cal H} (\xi) = \sum_{i=1}^3 (X_i -\xi_i {\bf 1}_{N\times N}) \otimes \sigma^i 
\end{equation}
A zero eigenvalue occurs at $\vec{\xi}$ exactly when $\det(\hat{\Hcal}(\xi)) = 0$. At this place, the determinant changes sign. We can therefore plot the level set $\det(\hat{\Hcal}(\xi))=0$ and standard numerical algorithms can be used to determine this locus. Since this is a polynomial equation in real variables, the corresponding surface is algebraic in nature. In contrast, for the BPS solutions at large $N$, we get a surface that also contains the $log$ function in equation \eqref{eq:curve}. In this sense, there is a measurable finite departure from the finite $N$ and the infinite $N$ limit. We can think of this procedure as measuring quantum geometry corrections to large $N$. The most glaring one is that the rotation symmetry group of the solution is reduced from $U(1)$ to $\BZ_N$, so the matrix solutions are lumpy, and the lumpiness is non-perturbative in $N$; we get exactly $N$ lumps.
   
Our goal stated at the beginning of the paper is to analyze the topology transition from a sphere to a torus. As we discussed in section \ref{sec:largeN}, at large $N$ the topology change is instantaneous as soon as we turn on a non-zero value of the angular momentum $\tilde \jmath$. As can be seen from figure \ref{fig:largeNt} for the BPS trajectory, this proceeds by forming a very thin funnel between the north pole and the south pole. This funnel represents a condensate of strings, as the picture of \cite{Nishioka:2008ib} suggests we should have. From here we ask a few natural questions. The first one is if the topology change is apparent as a phase transition in the bosonic set of degrees of freedom, that is, if we have to go beyond a phase transition in diagrams \ref{fig:phdiag2} , \ref{fig:4x4}, or  \ref{fig:fuzzy10}. The second question is if the transition is instantaneous or not. Lastly we ask what happens also when the torus gets very thin (as suggested by the thin ring solution in the continuum limit).

\begin{figure}[b]
\includegraphics[width=2.5 in]{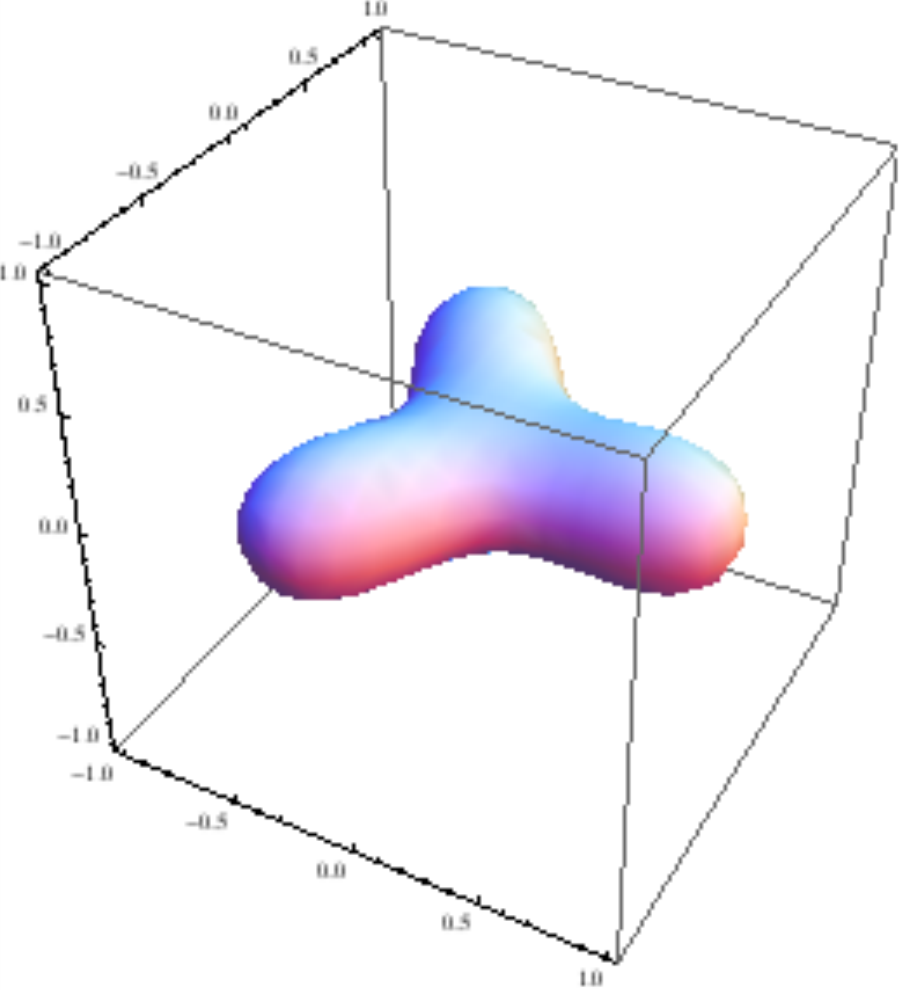}%
\includegraphics[width=2.5 in]{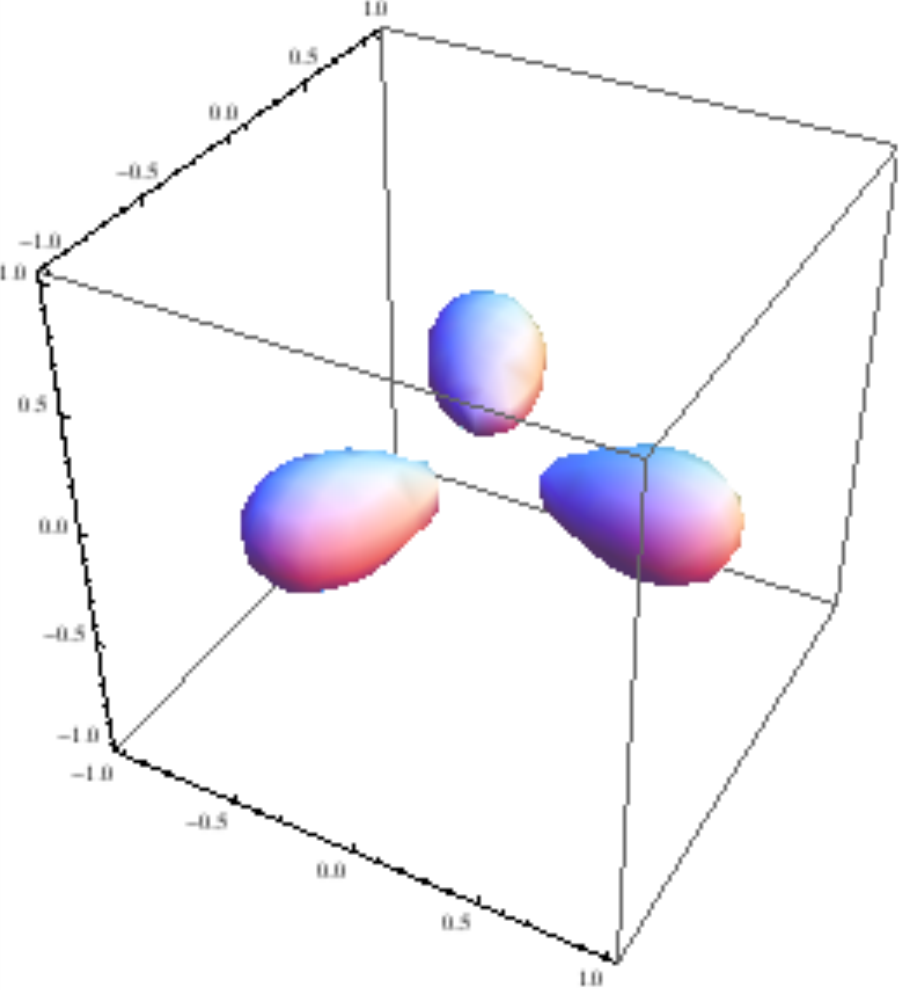}
\caption{Transition from a deformed sphere to three small spheres for the case of $3\times3$ matrices. The values are at $J\simeq 1.9$ and $J\simeq 1.99$ for the maximal sphere.}
\label{fig:3x3deg}
\end{figure}

We do not find tori for either $2\times 2$ or $3\times 3$ matrices. This is shown in figure \ref{fig:3x3deg}. The transition in topology is from a single sphere to three different small spheres. This can be understood as a transition from a membrane to a collection of separated D0-branes that have been puffed up a little bit from having some off-diagonal excitations, rather than pure D0-branes where the ansatz is diagonal. Incidentally, in the trivial configuration one can show that $Z$ commutes with $X^+$ and $X^-$ and they actually commute with each other, so this represents a collection of separated D0-branes, because the matrices can be diagonalized simultaneously. This is the same type of interpretation as in the BFSS matrix theory \cite{Banks:1996vh}.

\begin{figure}[t]
\includegraphics[width=2.5 in]{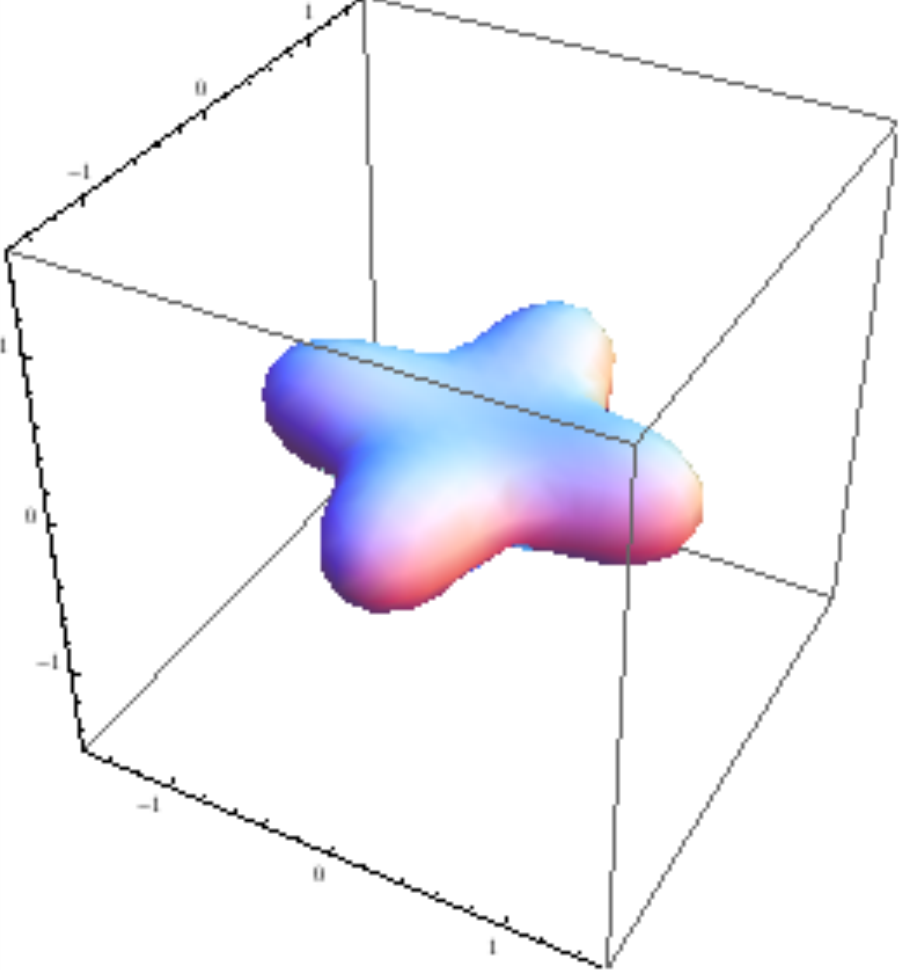}%
\includegraphics[width=2.5 in]{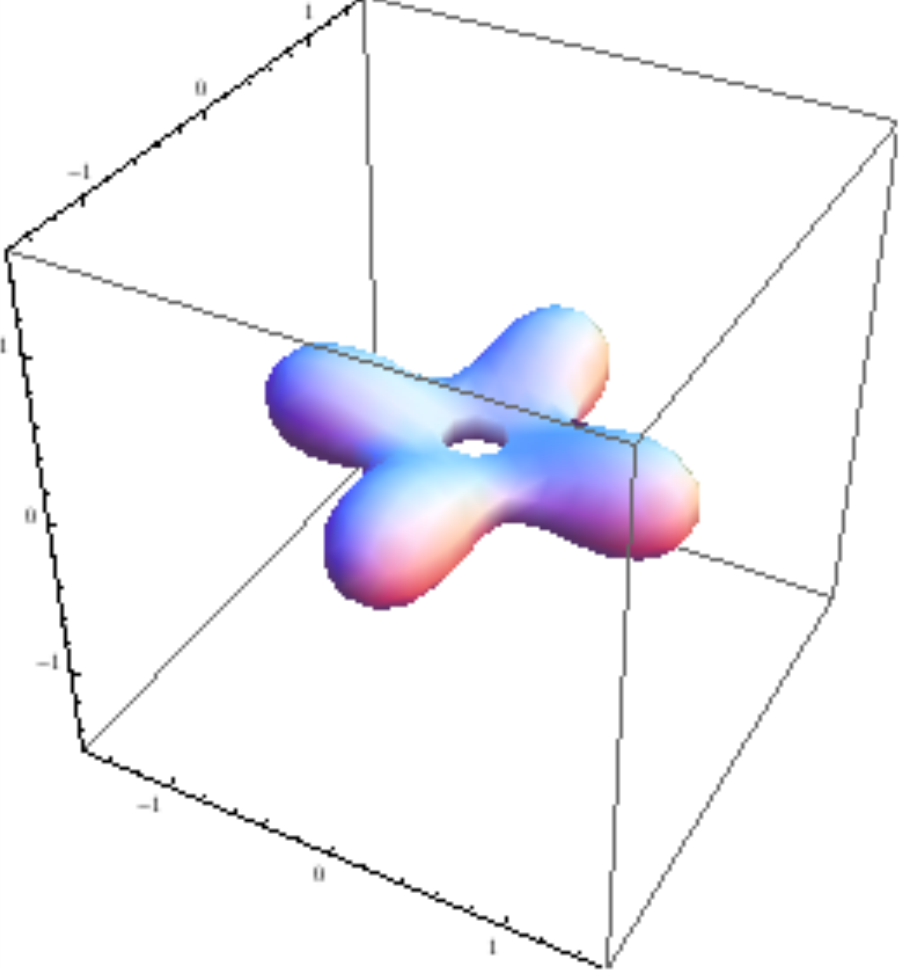}
\caption{Transition from a very distorted sphere to a very distorted tours  for $4\times4$ matrices. The values are at $J\simeq 4.24$ and $J\simeq 4.31$ for the maximal sphere.}
\label{fig:4x4deg}
\end{figure}

The topology transition depicted in figure \ref{fig:3x3deg} occurs before any singularity of the phase diagram in figure \ref{fig:phdiag1} is encountered, but for the values of $J$ given, it is close to the transition.
The same procedure for the case of $4\times 4$ matrices is depicted in figure \ref{fig:4x4deg} where we see a sphere transitioning to a torus near the phase transition for the maximal sphere, but before it. 

From these examples it should be clear that as far as the bosonic degrees of freedom are concerned, the change of topology from a sphere to a torus is smooth. Moreover, it does not occur immediately as in the large $N$ limit. The fermions do detect the topology change. The tori that are obtained this way for the maximal sphere for such small values of $N$ are very distorted. This should  improve as we increase $N$.

One could also analyze the geometry and topology in terms of the ideas of fuzzy Riemann surfaces found in \cite{Shimada:2003ks,Arnlind:2006ux}. For the analysis of topology, one uses properties of the eigenvalues of $Z$ interpreted as a Morse function. This only works for very large $N$.
As far as the geometry is concerned,  we find that the finite matrices for some sufficiently classical states (at sufficiently large $N$ again) would
give rise to fuzzy approximations to \eqref{eq:curve} for some values of $W$,  where $q\simeq (X^+X^-)$ is interpreted as a matrix and  is a normal ordered form of the product. 
These solutions are not algebro-geometric in nature because of the logarithm. In this case $q$ and $Z$ commute and can be thought of as classical variables on the torus that can be constrained by an equation. Deviations from satisfying the precise equation $W= z^2+q -1/4 \tilde\jmath \log(q)$ for fixed $W$ should then be interpreted as ``quantum corrections" in $1/N$. These are not quantum in the sense of due to a non-trivial vale of the Planck constant $\hbar$, but should be thought of instead of as quantum corrections to geometry due to the discrete nature of the D0-brane charge.  These corrections are  due to the small size of the M-theory circle in the DLCQ limit \cite{Seiberg:1997ad}. Here we take the DLCQ 
of the plane wave geometry. 

%%%%%%%%%%%%%%%%%%%%%%%%%%%%%%%%%%%%%%%%%%%%%%%%%%%%%%%%%%%%%%%%%%%%%%%%%%%%%%%

\section{Conclusion}

We have analyzed a particularly simple set of periodic classical solutions of the $SO(3)$ sector of the BMN matrix model. The solutions are periodic in time modulo gauge transformations. They are also required to preserve the maximal discrete subgroup of rotations along the axis of rotation that is allowed by the discreteness of the matrices, namely a $\BZ_N$. 
At large $N$, a $U(1)$ of rotations is recovered, so if the solutions have a continuous limit, they go to a rotationally invariant configuration. The solutions are not supersymmetric. As a consequence, the rotations that are turned on are not a central charge of the theory. There is, however, a BPS inequality that the solutions must satisfy, relating their energy to the angular momentum. Some solutions saturate the inequality, and they have a simpler set of equations that need to be satisfied. 

The rigidly rotating solutions can be understood in terms of a system of algebraic equations. These are in one to one correspondence with critical points of an energy function. Solutions found at zero angular momentum correspond not only to the vacua of the matrix model, made of concentric fuzzy spheres, but also other various unstable saddle points. These all survive when we turn on the angular momentum. The excitation of angular motion in the configurations always leads to a term in the on-shell energy proportional to the angular momentum, rather than starting at the square of angular momentum. This indicates that even though the configurations are rotating rigidly, the matrix object can not be thought of as a rigid body.

Both the finite $N$ and large $N$ results suggest that there is a maximal angular momentum beyond which there is only one phase. This should scale like $N^3$.
This phase when interpreted geometrically corresponds to the brane dissociating into a collection of D0-branes arranged symmetrically on a circle.

The phase diagram of solutions is rather complicated in general and we found it very useful to use Morse theory to find all the solutions, at least for low values of $N$. The full pattern for a given $N$ has many more saddles than the ones for smaller $N$. Any solution that is found for lower values of $N$ at $J=0$ can be combined with other such solutions to build solutions at a given $N$ for $J=0$, and these are seeds for a family of such solutions at finite $J$. Moreover, for every $N$ there are new solutions, some of which we know, like the maximal fuzzy sphere or a maximal fuzzy sphere at half radius.
We do not have a systematic way to search for the other ones.
The parameter space grows in dimension proportional to $N$ making it increasingly difficult to find them. Understanding this pattern in general should be very interesting.

At large $N$, the particular family of solutions we have considered reduce to a variational problem for critical points of an energy which is a local integral. The saddle point equations reduce to finding periodic solutions of a pair of coupled second order differential equations, while the BPS ones reduce to solving a coupled set of non-linear first order differential equations. It is only solutions with the right period that can be used. The period fixes the D0-brane charge of the configuration.

We also found that the topology change from a sphere to a (sometimes multiply wound) torus happens instantaneously in the large $N$ limit, but not so at finite $N$. Here it is delayed. Moreover, the shape of the fuzzy membranes can be very deformed from a circular torus. The effects that lead to that deformation are suppressed at large $N$. It should be interesting to investigate this in more detail. The breaking of the symmetry is due to the discretization of the D0-brane charge into  matrices. In the fuzzy geometry it is a purely classical effect. However, in the continuum large $N$ limit, this is supposed to arise from quantum effects that are responsible for the quantization of the D0-brane charge. It is often the case that quantum effects on D-brane field theories can be captured geometrically, like beta functions being captured by brane bending \cite{Randall:1998hc,Leigh:1998hj}. In this case we find that the discretization that appears in matrix theory introduces symmetry breaking effects that are not apparent in the continuum limit, and they would not appear in perturbation theory.

It should also be interesting to understand supersymmetry breaking effects better in these solutions and in particular the corrections due to zero point energy. This depends on the full theory, as we found that the $SO(3)$  BMN  system of classical equations arises in various different contexts. Such answers depend on the context.

Another interesting possibility to examine is that the $SO(3)$ sector of the BMN matrix model is also part of the description of the discrete lightcone quantization of the membrane in the Penrose limit of $AdS_4\times S^7$ \cite{Berenstein:2002jq}, or it's orbifolds. In particular, one can consider the ABJM model \cite{Aharony:2008ug} in the appropriate sector
that takes us to the Penrose limit. The natural candidates to consider are $D0$-brane states, which are dual to monopole operators. The spectrum of fluctuations around such objects  have been analyzed in \cite{Berenstein:2008dc, Berenstein:2009sa,Kim:2009ia,Kovacs:2013una} and they have many fluctuations that are supersymmetric and  saturate the BPS bound. It should be interesting to try to turn on rotations in $AdS_4$ for such states and see if the extremal solutions we have found can be mapped to them at weak coupling as well. This should help to understand how the D0-brane theory of the matrix model and the ABJM field theory in the presence of monopole states are related to each other.

%%%%%%%%%%%%%%%%%%%%%%%%%%%%%%%%%%%%%%%%%%%%%%%%%%%%%%%%%%%%%%%%%%%%%%%%%%%%%%%

\acknowledgments

D.B. would like to thank N. Obers for discussions. Work of D.B. supported in part by the U.S. Department of Energy under grant DE-SC0011702. 
The research leading to these results has received funding from the European Research Council under the European Community's Seventh Framework Programme (FP7/2007-2013) / ERC grant agreement no. [247252]. E. D. supported  by the Department of Energy Office of Science Graduate Fellowship Program (DOE SCGF), made possible in part by the American Recovery and Reinvestment Act of 2009, administered by ORISE-ORAU under contract no. DE-AC05-06OR23100. 

%%%%%%%%%%%%%%%%%%%%%%%%%%%%%%%%%%%%%%%%%%%%%%%%%%%%%%%%%%%%%%%%%%%%%%%%%%%%%%%

\end{document}